\documentclass[11pt,a4paper]{article}
\usepackage{heppub}
\usepackage[english]{babel}
\usepackage{amssymb,amsmath}
\usepackage{mathrsfs}
\usepackage{hyperref}
\usepackage[dvipsnames,x11names]{xcolor}
\usepackage{graphicx}
\usepackage{subfig}
\usepackage{mathtools}
\usepackage{slashed}
\usepackage{float}
\usepackage[normalem]{ulem}
\usepackage{accents}

\makeatletter
\gdef\@fpheader{}
\makeatother

\usepackage{dcolumn}
\usepackage{textcomp}
\usepackage[dvipsnames]{xcolor}
\usepackage{xfrac}
\usepackage{slashed}
\usepackage{multirow}

\usepackage{cleveref}
\usepackage{tikz}
\newcommand*\circled[1]{\tikz[baseline=(char.base)]{\node[shape=circle,draw,inner sep=2pt] (char) {#1};}}
\def\be{\begin{equation}}
\def\ee{\end{equation}}
\def\figs/B{B}
\def\bea{\begin{eqnarray}}
\def\eea{\end{eqnarray}}
\def\bg{\begin{eqnarray}}
\def\nd{\end{eqnarray}}

\def\log{{\rm log}}

\usetikzlibrary{decorations.pathmorphing}
\tikzset{snake it/.style={decorate, decoration=snake}}

\def\beq{\begin{equation}}
\def\eeq{\end{equation}}

\newcommand{\intd}{\int {\rm d}}

\renewcommand{\d}{{\rm d}}

\definecolor{oiOrange}{RGB}{230,159,0}
\definecolor{oiBlue}{RGB}{0,114,178}
\definecolor{oiGreen}{RGB}{0,158,115}

\newcommand{\sint}[3]{\hspace{- #3 em} \underset{#1}{\overset{#2}{\int}}\hspace{- #3 em}}

\preprint{MIT-CTP/5885 \begin{flushright}IFT-UAM/CSIC-25-145 \end{flushright}}

\title{Quantum tunneling from excited states in the steadyon picture} 

\author[a,b]{Joshua Lin,}
\author[c]{Bruno Scheihing-Hitschfeld,}
\author[a,d,e]{and Thomas Steingasser}

\affiliation[a]{%
MIT Center for Theoretical Physics -- a Leinweber Institute, Massachusetts Institute of Technology, Cambridge, MA  02139, USA}

\affiliation[b]{Physics Division, Argonne National Laboratory, Lemont, IL 60439, USA}

\affiliation[c]{%
Kavli Institute for Theoretical Physics, University of California, Santa Barbara, California 93106, USA
}

\affiliation[d]{Black Hole Initiative at Harvard University, 20 Garden Street, Cambridge, MA 02138, USA
}

\affiliation[e]{Departamento de Fisica Teorica, Universidad Autonoma de Madrid, and IFT-UAM/CSIC, Cantoblanco, 28049, Madrid, Spain
}

\emailAdd{joshua.lin@anl.gov}
\emailAdd{bscheihi@kitp.ucsb.edu}
\emailAdd{thomas.steingasser@uam.es}

\abstract{Recent developments in the understanding of real-time path integrals led to the development of the ``steadyon picture'' for the semi-classical calculation of quantum tunneling rates. We discuss tunneling out of a generic localized initial state in this picture and present its application for the important example of a resonance state in a one-dimensional point particle potential. We find that the steadyon picture indeed reproduces existing results obtained using the WKB method. Our analysis furthermore demonstrates how applying this picture to physical states naturally addresses open conceptual questions regarding this framework. Finally, we perform a numerical study for a specific potential. We demonstrate in particular the existence of regimes in which the tunneling rate is dominated by higher resonances, rather than the false vacuum, as well as their importance. 
}

\begin{document}
\maketitle

\section{Introduction}
After having served as one of the most emblematic features of all quantum theories for nearly a century~\cite{Wentzel:1926WKB,Hund:1927Tunnel,Gamow:1928AlphaDecay,GurneyCondon:1928Nature,GurneyCondon:1929PR,MacColl:1932WavePackets,Coleman:1977py,Callan:1977pt,Kobzarev:1974cp,Devoto:2022qen}, quantum tunneling still remains a subject of active research. On a formal level, one of the biggest challenges remains the development of a first-principles based formalism capable of providing an analytical understanding of tunneling out of initial states other than the well-understood false vacuum~\cite{Steingasser:2024ikl,Steingasser:2023gde}.

For the calculation of tunneling rates out of a false vacuum, the Euclidean-time framework first proposed in Ref.~\cite{Coleman:1977py} and recently improved upon through the introduction of the ``direct approach'' in Refs.~\cite{Andreassen:2016cff,Andreassen:2016cvx} has proven both insightful and practical. However, as pointed out in Ref.~\cite{Steingasser:2024ikl}, the Wick rotation underlying this approach becomes problematic for essentially any initial state other than a false vacuum state. The authors of Ref.~\cite{Steingasser:2024ikl} therefore suggested to instead evaluate the path integrals in real time, ensuring the existence of an analytical saddle point through an infinitesimal deformation of the Hamiltonian,
\begin{align}
	H \to (1 - i \epsilon )H \, .
\end{align}
These authors then showed that for a properly chosen deformation parameter $\epsilon$ and a physical time significantly larger than the typical time scale of the system, theories deformed in this manner give rise to complex saddle point solutions dubbed \textit{steadyons}. For more general investigations of such complex solutions, consider Refs.~\cite{Witten:2010cx,Tanizaki:2014xba,Cherman:2014sba,Dunne:2015eaa,Bramberger:2016yog,Michel:2019nwa,Mou:2019gyl,Hertzberg:2019wgx,Ai:2019fri,Hayashi:2021kro,Nishimura:2023dky}. Crucially, the tunneling rate $\Gamma$ at any physical time $t$ obtained using these solutions can, in the appropriate limits, be expressed through the Euclidean action of an instanton $x_{I} [\tau]$ defined on the Euclidean-time interval $\Delta \tau \simeq  \epsilon t$,
\begin{align}
	\Gamma \sim e^{-2 S_E [x_I]} \sim e^{-S_E [x_b]}\, ,
\end{align}
where $x_b [\tau]$ denotes the bounce solution corresponding to $x_I [\tau]$. Throughout this article, we denote by \textit{instanton} a Euclidean-time solution traversing the potential barrier protecting the false vacuum once, while we use \textit{bounce} for a solution that subsequently returns to it starting point.

On a first look, this would seem to suggest that the result for $\Gamma$ should ultimately depend on the chosen deformation. While this could indeed present a possible problem, it is natural to expect that this apparent ambiguity must be resolved when actual physical states are being considered to unambiguously determine the problem's boundary conditions.\footnote{Note that Ref.~\cite{Steingasser:2024ikl}, due to its focus on the properties of the steadyon solutions, considers only the unphysical example of an exact position eigenstate. For this special case, the rate does indeed depend on the chosen value of $\epsilon$. These authors propose to interpret this observation as each choice of $\epsilon$ selecting a different momentum mode forming the position eigenstate, suggesting to sum over all values of $\epsilon$ -- or, equivalently, $\Delta \tau$. In this article, we will find that the tunneling process can indeed be understood as unfolding through the sum over one-parameter family of instantons, characterised by different Euclidean-time intervals $\Delta \tau$. We will, however, find that for a physical eigenstate the result becomes independent of the choice of $\epsilon$.} This conjecture is confirmed by the results of Ref.~\cite{Steingasser:2023gde}, where the steadyon picture was used to analytically derive the tunneling rate out of a supercooled system at temperature $T\equiv \beta^{-1}$. These authors then argue that the density matrix representing such a system, $\rho \sim \exp(- \beta H)$ (restricted to the false vacuum basin) can for large enough physical times $t$ be understood as a Wick rotation by some finite angle $\epsilon \sim \beta/t$ by deforming the Keldysh-Schwinger contour describing the tunneling process. Thus, for this example the properties of the initial state indeed naturally provide the deformation parameter, ultimately allowing for the first completely first-principles based derivation of the finite-temperature tunneling rate.

The main goal of this work is the extension of this picture (and the underlying ``direct approach'') to allow for the description of generic point particle tunneling processes out of some basin $\mathcal{F}$ in any potential where tunneling can happen. Our first main result is an expression for the rate of probability flux out of $\mathcal{F}$ using the direct approach, applicable to any initial state that can be represented through a wave function $\psi$. We then demonstrate explicitly how this expression can be evaluated in the steadyon picture, considering for concreteness the case where the initial state may be decomposed in terms of ``resonance states.'' This not only represents another major step towards the calculation of tunneling rates in increasingly complicated systems, but also serves to improve the understanding of the steadyon framework and its conceptual subtleties.

This article is structured as follows. First, in Sec.~\ref{sec:tunnel-steady}, we use the direct approach to derive a closed expression for the probability flux for a point particle tunneling out of a basin in an arbitrary potential as a function of its wave function. Using the steadyon picture, we are then able to evaluate the parts of this expression that are independent of the wave function to leading order in the semi-classical expansion. This reduces the task of evaluating the probability flux to the search of a saddle point subject to boundary conditions set by the initial state, which can be obtained from our formula in a straightforward way. In Sec.~\ref{sec:resonance}, we use these properties to analytically evaluate the expression for the probability flux out of resonance states in the steadyon picture, from which it is straightforward to construct the corresponding tunneling rate. We confirm that our result agrees with the existing literature using the WKB approximation. As these states serve as a (non-orthogonal) basis for states localised within the basin, this gives us the ability to fully analyse the tunneling behavior of an arbitrary state, given that its overlap with these resonance states can be obtained. In Sec. 4 we perform a numeric investigation of a specific 1D-potential, preparing a number of initial wavefunctions and fitting the resulting tunneling rates to the expectation given a discrete resonant expansion. As expected, at large physical times the tunneling rate $\Gamma$ approaches a constant given approximately by the expressions derived in Sec. 2. At intermediate times $\Gamma$ is shown to have oscillations which can be perfectly described by mixing between the non-orthogonal resonant states. We furthermore investigate initial states that dominantly overlap onto excited resonant states, leading to a `multi-slope' behavior where the dominant contribution to $\Gamma$ arises from different resonant states at different times. To keep this article as self-contained as possible, we briefly review the steadyon picture in Appendix~\ref{app:steady-review}. 

\section{Tunneling in the steadyon picture} \label{sec:tunnel-steady}

\begin{figure}
\centering
\begin{tikzpicture}[scale=5]
  \draw[->] (0,0) -- (1.15,0) node[right] {$x$};
  \draw[->] (0,0) -- (0,0.4);
  \draw (0.81,0.41) node [black]  {$V(x)$}; 
\draw[domain=0:1.079,smooth,variable=\x,very thick] plot ({\x},{0.8*(\x)^2 - 0.6*(\x)^6});  
\draw[gray!60,blue,snake it, thick] (0,0.2)--(0.51,0.2);
  \draw[gray!60,dashed] (0.51,0)--(0.51,0.2);
  \draw[gray!60,dashed] (1,0)--(1,0.2);
  \draw (-0.01,-0.05) node [black] {$x_{\rm FV}$};
  \draw (0.51,-0.05) node [black]  {$x_*$};
  \draw (1,-0.05) node [black]  {$x_s$};
  \draw (0.3,0.4) node [black]  {$\mathcal{F}$};
  \draw (1.1,0.4) node [black]  {$\mathcal{R}$};
  \node at (0.51,0.2) [circle,draw=gray,fill=gray!60]{};
\end{tikzpicture}
\begin{tikzpicture}[scale=5]
  \draw[->] (0,0) -- (1.15,0) node[right] {$x$};
  \draw[->] (0,0) -- (0,0.4);
  \draw (0.81,0.41) node [black]  {$V(x)$}; 
\draw[domain=0:1.079,smooth,variable=\x,very thick] plot ({\x},{0.8*(\x)^2 - 0.6*(\x)^6});  
\draw[gray!60,blue,snake it, thick] (0,0.2)--(0.51,0.2);
\draw[domain=0.51:1.06,smooth,variable=\x, thick,blue] plot ({\x},{0.055/\x^2}); 
\draw[gray!60,blue,snake it, thick] (1.06,0.05)--(1.15,0.05);
  \draw[gray!60,dashed] (0.51,0)--(0.51,0.2);
  \draw[gray!60,dashed] (1,0)--(1,0.2);
  \draw (-0.01,-0.05) node [black] {$x_{\rm FV}$};
  \draw (0.51,-0.05) node [black]  {$x_*$};
  \draw (1,-0.05) node [black]  {$x_s$};
  \draw (0.3,0.4) node [black]  {$\mathcal{F}$};
  \draw (1.1,0.4) node [black]  {$\mathcal{R}$};
  \node at (0.51,0.2) [circle,draw=gray,fill=gray!60]{};
  \node at (1,0.2) [circle,draw=gray,fill=gray!60]{};
  \draw[->,gray!60,very thick] (0.58,0.2)--(1,0.2);
  \draw[->,gray!60,very thick] (0.58,0.2)--(0.84,0.2);
  \draw[->,gray!60,very thick] (0.51,0.2)--(0.62,0.2);
\end{tikzpicture}
\caption{Quantum tunneling of a point particle in a potential (black), represented by its wave function (blue). \textit{Left panel}: At $t=0$, the particle is fully localised within the well $\mathcal{F}$. $x_*$ denotes the point beyond which its wave function decays exponentially. \textit{Right panel}: At times larger or comparable to the potential's typical time scale $t \gtrsim t_{\rm sys}$, the particle's wave function has penetrated the potential barrier separating $\mathcal{F}$ from the adjacent basin $\mathcal{R}$ into which the particle tunnels. We assume that the exponential decay of the wave function $x_*$ remains true for long enough times to allow for ``steady'' tunneling. If the point beyond which the exponential decay of the wave function sets off varies with time, we call $x_*$
the largest value during the considered time interval. $x_s$ denotes the point on the other side of the barrier with $V(x_s)=V(x_*)$.}
\label{fig:FRBarrier}
\end{figure}
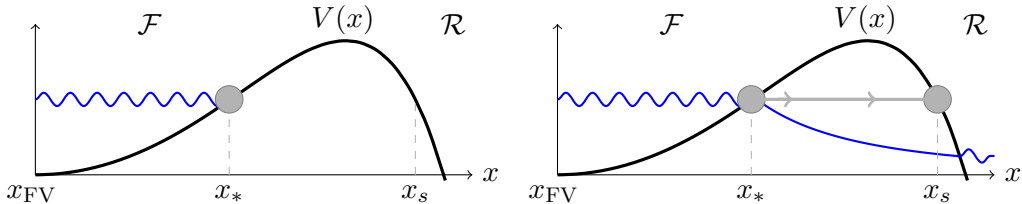

Following the direct approach developed in Refs.~\cite{Andreassen:2016cff,Andreassen:2016cvx}, we describe the tunneling process in terms of the probability to find the particle in different regions of space. For simplicity, we restrict ourselves to the case of a point particle in one spatial dimension described by a wave function $\psi(x)$. At $t=0$, the particle is entirely localised within some basin $\mathcal{F}$. Furthermore, we assume that the particle can only escape in one direction and we denote the region behind the potential barrier in this direction as $\mathcal{R}$. See Fig.~\ref{fig:FRBarrier}.

This allows us to define the probability to find the particle in $\mathcal{F}$ or $\mathcal{R}$, respectively, at some physical time $t$ as
\begin{align}
    P_{\Omega} (t) =& \int_{\Omega} \d x\ \langle x | e^{-i H t} | \Psi \rangle \langle \Psi | e^{i H t}|x\rangle \nonumber \\ 
    =& \int_{\Omega} \d x \intd x_1 \intd x_2\ \psi(x_1) \psi^* (x_2) \langle x | e^{-i H t} | x_1 \rangle \langle x_2 | e^{i H t}|x\rangle \nonumber \\
    \equiv& \intd x_1 \intd x_2\ \psi(x_1) \psi^* (x_2)  p_{\Omega} (x_1,x_2,t)\label{eq:Pgen} \, ,
\end{align}
where $\Omega\in \{\mathcal{F},\mathcal{R} \}$. Here, as throughout the remainder of this article, any time-dependent quantity without explicit time label is supposed to be understood as evaluated at the initial time $t=0$. 

If $\mathcal{F}$ is itself localised around some false vacuum, it is easy to see that for any point $x_*$ defining the boundary of $\mathcal{F}$ there exists an energetically degenerate point on the other side of the barrier, $x_s \in \mathcal{R}$, beyond which an initial wave function prepared in $\mathcal{F}$ will be able to propagate freely.\footnote{As we will discuss later, for a resonance state, $x_*$ would be given by the classical turning point of a particle with the same energy as this state inside the false vacuum basin.}

Moving forward, we will find it convenient to note that the function $p_{\Omega} (x_1,x_2,t)$ is just a product of real-time propagators,
\begin{align}
    p_{\Omega} (x_1,x_2,t)=\langle x | e^{-i H t} | x_1 \rangle \langle x_2 | e^{i H t}|x\rangle = D_F (x,t|x_1) D_F^* (x,t|x_2) \, ,
\end{align}
which, to be completely explicit, are defined as
\begin{equation}
     D_F (x_2,t_2|x_1,t_1) \equiv \langle x_2| e^{-iH (t_2 - t_1) } | x_1 \rangle \, . 
\end{equation}
Using these probabilities, it is now straightforward to define the \textit{tunneling rate} $\Gamma(t)$ of the initial state through the relation
\begin{align}
	P_{\mathcal{F}}(t) = P_{\mathcal{F}}(0) \cdot e^{- \Gamma(t) t} \, .
\end{align}
This definition becomes useful in cases in which $\Gamma(t)$ is approximately constant during the tunneling process, as is the case for tunneling out of a false vacuum, resonance or supercooled state.\footnote{As pointed out in Refs.~\cite{Andreassen:2016cff,Andreassen:2016cvx}, this behavior can be expected only for times larger than the typical timescale of the system, but shorter than the typical time of tunneling. We revisit this matter in Sec.~\ref{sec:numbers}.} In these cases, it is straightforward to relate $\Gamma$ to the probabilities $P_{\mathcal{F}},P_{\mathcal{R}}$ through~\cite{Andreassen:2016cff,Andreassen:2016cvx}
\begin{align}
	\Gamma(t) = - \frac{\dot{P}_{\mathcal{F}} (t)}{P_{\mathcal{F}} (t)}= \frac{\dot{P}_{\mathcal{R}} (t)}{P_{\mathcal{F}} (t)} \, .
\end{align}
We will begin by analyzing the probability flux $\dot{P}_{\mathcal{R}} (t)$, and evaluate $P_{\mathcal{F}} (t)$ directly for the resonance states of interest in Sec.~\ref{sec:resonance}.

In order to calculate $\dot{P}_{\mathcal{R}} (t)$, we can first observe that Eq.~\eqref{eq:Pgen} allows us to separate the initial condition, encoded in the wave function $\psi (x)$, from the dynamics encoded in the function~$p_{\mathcal{R}} (x_1,x_2,t)$. Next, following Refs.~\cite{Andreassen:2016cff,Andreassen:2016cvx,Steingasser:2024ikl,Steingasser:2023gde}, we can decompose the propagators connecting a point $x_i$ in the basin $\mathcal{F}$ with a point $x_f$ in $\mathcal{R}$ as
\begin{align}
	D_F(x_f,t | x_i)= \int_{0}^t \d t_s D_F( x_f, t |x_s, t_s ) \ \bar{D}_F(x_s,t_s|x_i) \, ,
\label{eq:Ddecomp}
\end{align}
where $\bar{D}_F$ is defined through
\begin{equation}
    \bar{D}_F (x_s , t_s | x_i) \equiv 
    \sint{x (0) = x_i}{x (t_s) = x_s}{1} {\mathcal D} x \ e^{i S[x]} \delta \left( F_{x_s}[x] - t_s \right) \, . \label{eq:Dbar}
\end{equation}
The functional $F_{x_s}$ is defined as mapping any path onto the time it first crosses the \textit{emergence point} $x_s$. As written here, $x_s$ can be arbitrary, but as we anticipated earlier, we will use it (along with its energetically degenerate counterpart $x_*$) to define the region we call the ``barrier'', $\{x : x_* \leq x \leq x_s\}$. We note that this only becomes precise once one specifies the state, as we do in Sec.~\ref{sec:resonance}.

This decomposition has a simple intuitive interpretation. In order for the particle to move from $x_i$ into $\mathcal{R}$, it needs to cross $x_s$. As the path integral contains a sum over all possible paths, this could happen for the first time at any time $t_s\leq t$, represented by the integral over $t_s$. The condition that this occurs for the first time at the physical time $t$ was dubbed ``crossing condition'' in Ref.~\cite{Steingasser:2024ikl}. However, anticipating later steps of this discussion, we will specify this to \textit{final crossing condition}.

This decomposition now allows us to rewrite~$p_{\mathcal{R}} (x_1,x_2,t)$ as
\begin{align}
    p_{\mathcal{R}}(x_1,x_2,t)  = & \int_\mathcal{R}\d x_f\ D_F(x_{f},t|x_1)D_F^*(x_{f},t|x_2) \nonumber \\ 
    =& \int_\mathcal{R}\d x_f \int_0^t \d t_s  \int_0^t \d t_s^\prime \ D_F ( x_f , t | x_s , t_s) \bar{D}_F (x_s ,t_s | x_1) \nonumber \\
    & \qquad\qquad\qquad\qquad\qquad \times D_F^* (  x_f , t | x_s , t_s^\prime )  \bar{D}_F^* ( x_s ,t_s^\prime | x_2 ) \, .
\end{align}
In order to further simplify this, we need to make one last assumption, namely that in the regime of interest for us, back-tunneling is negligible. In terms of propagators, this amounts to neglecting paths starting in $x_s$ and ending in $\mathcal{F}$ (or inside the potential barrier), which can be formalised as
\begin{align}
    \int_\mathcal{R}\d x_f\  D_F (  x_f , t | x_s , t_s )  D_F^* (  x_f , t | x_{s} , t_s^\prime ) &\simeq  \int \d x_f\  D_F (  x_f , t | x_s , t_s )  D_F^* (  x_f , t | x_{s} , t_s^\prime ) \nonumber \\ &= D_F ( x_s , t_s^\prime | x_s , t_s  ) \, .
\end{align}
Next, we can rearrange the time integrals using the standard identity
\begin{equation}
    \int_0^t \text{d}t_s  \int_0^t \text{d}t_s^{\prime} = \int_0^t \text{d}t_s  \int_0^{t_s} \text{d}t_{s^{\prime}} + \int_0^t \text{d}t_s^{\prime} \int_0^{t_s^{\prime}} \text{d}t_s  \, .
\end{equation}
And indeed, taking the time derivative of $p_{\mathcal{R}}(x_1,x_2,t)$ has now become trivial,
\begin{align}
    \dot{p}_{\mathcal{R}}(x_1,x_2,t)  &=  \frac{\d}{\d t} \int_0^t \d t_s \ \int_0^t \d t_s^\prime \ \bar{D}_F ( x_s ,t_s | x_1 ) \bar{D}_F^* ( x_s ,t_s^\prime | x_2 )  D_F (  x_s , t_s^\prime | x_s , t_s  ) \nonumber \\ 
    &=  \int_0^t \d t_s \ \bar{D}_F ( x_s ,t_s | x_1 ) \bar{D}_F^* ( x_s ,t | x_2 )   D_F ( x_s , t  | x_s , t_s ) \nonumber \\ 
    & \quad +\int_0^t \d t_s^\prime \ \bar{D}_F ( x_s ,t | x_1 ) \bar{D}_F^* ( x_s ,t_s^\prime | x_2 )  D_F ( x_s , t_s^\prime | x_s , t )  \nonumber \\ 
    &=   D_F (  x_s , t | x_1 ) \bar{D}_F^* ( x_s ,t | x_2 )  +  \bar{D}_F ( x_s ,t | x_1 ) D_F^* (   x_s , t | x_2  ) \, .
\end{align}
We can now recall that we will need to integrate this quantity against the initial wave function $\psi (x_{1,2})$, and observe that the position of the argument $x_{1,2}$ inside the propagators forming $\dot{p}_{\mathcal{R}}(x_1,x_2,t) $ suggests a way to combine these quantities. In order to do so, we can now first introduce a second functional $G_{x_*}[x]$. Similar to the previously defined functional $F_{x_s}$, this functional maps any finite path onto the time it \textit{last} crosses $x_*$. This functional now allows us to define two more modified propagators as
\begin{align}
    \underaccent{\bar}{D}_F ( x_s,t_s | x_* ,t_*) \equiv & \quad \sint{x(t_*) = x_*}{x(t_s) = x_s}{1} \mathcal{D}x \ e^{iS[x]} \delta (G_{x_*} [x] -t_*) \, , \label{eq:Dunderbar} \\ 
    \bar{\underaccent{\bar}{D}}_F ( x_s,t_s | x_* ,t_* )   \equiv & \quad \sint{x(t_*) = x_*}{x(t_s) = x_s}{1} \mathcal{D}x \ e^{iS[x]} \delta (G_{x_*} [x] -t_*)  \delta (F_{x_s} [x] -t_s)  \, . \label{eq:Dbarunderbar}
\end{align}
These quantities allow us to decompose the already introduced propagators as
\begin{align}
    D_F (  x_s , t | x_i )= & \int_0^t \d t_* \  \underaccent{\bar}{D}_F (x_s,t | x_* ,t_*) D_F(x_*,t_* | x_i) \, , \\
    \bar{D}_F ( x_s , t | x_i )= & \int_0^t \d t_* \  \bar{\underaccent{\bar}{D}}_F ( x_s,t | x_* ,t_*) D_F(x_*,t_* | x_i) \, .
\end{align}
This decomposition is not only formally similar to the previous one, but also has a similar interpretation. Due to the imposed localisation of the wave function, the points $x_s$ and $x_*$ are defined such that any path starting at some $x_i$ where the initial wave function has support needs to cross $x_*$ before reaching $x_s$ -- and only beyond the latter the particle is considered to have successfully crossed the barrier.\footnote{In Sec.~\ref{sec:perstead}, we demonstrate the connection between $x_*$, $x_s$ and the initial state explicitly for the example of a resonance state.}

Using these propagators and the decomposition for $\dot{p}_{\mathcal{R}}$ shown above one obtains
\begin{align}
    \dot{P}_{\mathcal{R}} = & \intd x_1 \intd x_2 \ \psi(x_1) \psi^* (x_2) \dot{p}_{\mathcal{R}} (x_1,x_2,t) \nonumber \\ 
    =& \intd x_1 \intd x_2 \ \psi(x_1) \psi^* (x_2)\left[ D_F (  x_s , t | x_1 ) \bar{D}_F^* ( x_s ,t | x_2 )  +  \bar{D}_F ( x_s ,t | x_1 ) D_F^* (  x_s , t | x_2  ) \right] \nonumber \\ 
    =&\intd x_1 \intd x_2 \ \psi(x_1) \psi^* (x_2) \times  \nonumber \\ 
    & \bigg[ \int_0^t \d t_* \ \underaccent{\bar}{D}_F (x_s,t | x_* ,t_*) D_F(x_*,t_*|x_1)  \int_0^t \d t_*^\prime \ \bar{\underaccent{\bar}{D}}_F^* (x_s,t | x_* ,t_*^\prime)  D_F^*(x_*,t_*^\prime | x_2)   \nonumber \\ 
    &+ \int_0^t \d t_* \ \bar{\underaccent{\bar}{D}}_F ( x_s,t | x_* ,t_*) D_F(x_*,t_* | x_1)  \int_0^t \d t_*^\prime \ \underaccent{\bar}{D}_F^* (x_s,t | x_* ,t_*^\prime) D_F^* (x_*,t_*^\prime | x_2) \bigg] \, ,
\end{align}
and then it is straightforward to perform the integral over the initial wave function by noting that this is nothing more than the time-evolved wavefunction
\begin{align}\label{eq:propInt}
    \intd x_1 D_F(x_*,t_*|x_1) \psi(x_1) = \psi (x_*, t_*) \, ,
\end{align}
where the integral over $x_1$ may be restricted to $\mathcal{F}$ due to the support of the initial condition being localised it this region. Then, we conclude that
\begin{align}
    \dot{P}_{\mathcal{R}} = & \int_0^t \d t_* \int_0^t \d t_*^\prime \ \psi (x_*,t_*) \psi^* (x_*,t_*^\prime) \underaccent{\bar}{D}_F (x_s,t | x_* ,t_*) \bar{\underaccent{\bar}{D}}_F^* (x_s,t | x_* ,t_*^\prime) \nonumber \\ 
    &+ \int_0^t \d t_* \int_0^t \d t_*^\prime \  \psi (x_*,t_*) \psi^* (x_*,t_*^\prime) \bar{\underaccent{\bar}{D}}_F ( x_s,t | x_* ,t_*) \underaccent{\bar}{D}_F^* (x_s,t | x_* ,t_*^\prime) \nonumber \\ 
    =& \int_0^t \d t_* \int_0^t \d t_*^\prime \ \psi (x_*,t_*) \psi^* (x_*,t_*^\prime) \underaccent{\bar}{D}_F (x_s,t | x_* ,t_*) \bar{\underaccent{\bar}{D}}_F^* (x_s,t | x_* ,t_*^\prime) +c.c. \nonumber \\  
    =&\int_0^t \d \Delta t  \ \underaccent{\bar}{D}_F (x_s,\Delta t | x_* ) \psi (x_*,t-\Delta t)  \int_0^t \d \Delta t^\prime \ \bar{\underaccent{\bar}{D}}_F^* (x_s, \Delta t^\prime | x_* ) \psi^* (x_*,t-\Delta t^\prime)  +c.c. \, .\label{eq:P-flux-master}
\end{align}
Thus, we have reduced the calculation of the probability flux to the evaluation of this integral, mixing factors representing the time evolution of the wave function within $\mathcal{F}$ with propagators describing the tunneling.\footnote{Insofar as processes that tunnel to $\mathcal{R}$ and back can be neglected.} To keep our discussion as general possible for as long as possible, we can now first focus on the propagators, which can be evaluated through the steadyon method reviewed in Appendix~\ref{app:steady-review}. 

Following Ref.~\cite{Steingasser:2024ikl}, we regularise the theory by introducing a small deformation of the Hamiltonian as
\begin{align}\label{eq:regm}
    H \to (1 - i \epsilon) H \, ,
\end{align}
with \textit{some} small $\epsilon$ and assume that $t\gg t_{\rm sys}/\epsilon$, where $t_{\rm sys}$ denotes the system's typical time scale. Next, we can denote by $\Delta \tau_{\rm per}$ denote the Euclidean time it takes a particle setting out from rest at $x_*$ to reach the point $x_s$ \textit{in Euclidean time}, as well as $\Delta t_{\rm per} \equiv \epsilon\cdot \Delta \tau_{\rm per} $. Then, we recall that a steadyon connecting $x_*$ and $x_s$ while in the limit $\epsilon \to 0$ satisfying the boundary and crossing conditions exists if $\epsilon \cdot \Delta t\simeq  \Delta \tau$, where $\Delta \tau< \Delta \tau_{\rm per}$ -- and thus, $\Delta t<\Delta t_{\rm per}$.\footnote{Note that, in principle, the steadyon is only capable of matching the boundary condition for a countable number of times, $t= N \cdot \Delta T$, where $\Delta T$ is the classical oscillation time of the particle inside $\mathcal{F}$. For a smooth wave function, however, this can be circumvented easily by noting that in the limit $t\to \infty$ an infinitesimal change in $x_*$ will be grown to a large enough offset at $t$ to compensate for this.} This bound is easiest understood by considering the projection of the steadyon onto the Euclidean-time axis. For a particle to move from $x_*$ to $x_s$ in an Euclidean time larger than $ \Delta \tau_{\rm per}$, it would either have to start out with an initial velocity away from $x_s$ or with a large enough positive velocity to ``overshoot'' $x_s$, turn around and then move back. These options would, however, violate the initial and final crossing condition in $\bar{\underaccent{\bar}{D}}_F ( x_s,\Delta t | x_*)$, respectively.\footnote{Another subtlety of this picture is the definition of the crossing conditions for a complex solution. In principle, any finite $\epsilon$ will lead to an imaginary part of $\mathcal{O}(\epsilon)$ near $t_s$. Thus, we define the crossing conditions as the path hitting the emergence point \textit{up to} an error $|x(t)-x_s|_{\mathbb{C}}\sim \mathcal O(\epsilon)$.} The propagator $\underaccent{\bar}{D}_F ( x_s,\Delta t^\prime | x_*)$, meanwhile, is also subject to the initial crossing condition, but not its final counterpart. Thus, solutions where the steadyon (or, easier to visualise, its projection on the Euclidean-time axis) overshoot the emergence point and spend some time ``hovering'' near the true vacuum are in principle suitable saddle points for the evaluation of the propagator $\underaccent{\bar}{D}_F ( x_s,\Delta t | x_*) $. It is, however, also easy to see that such saddle points will lead to an exponentially larger suppression than their counterparts arriving at $x_s$ directly. The complex conjugate of this term behaves identical under the replacement $\Delta t \leftrightarrow \Delta t^\prime$.

For any pair $\{\Delta t, \Delta t^\prime \}$, there exists a pair of steadyon solutions that can be used to evaluate the path integrals in a stationary phase approximation. However, as the total flux is obtained by integrating over all possible $\Delta t$ and $\Delta t^\prime$, we find that the flux is indeed controlled by a sum over all of these steadyons. In Euclidean-time language, the tunneling unfolds through a superposition of a one-parameter family of contributions represented by instantons with every possible $\Delta \tau$. Crucially, this implies that, as long as $\epsilon< \Delta \tau_{\rm per}/t$, our result is ultimately \textit{independent} of $\epsilon$.

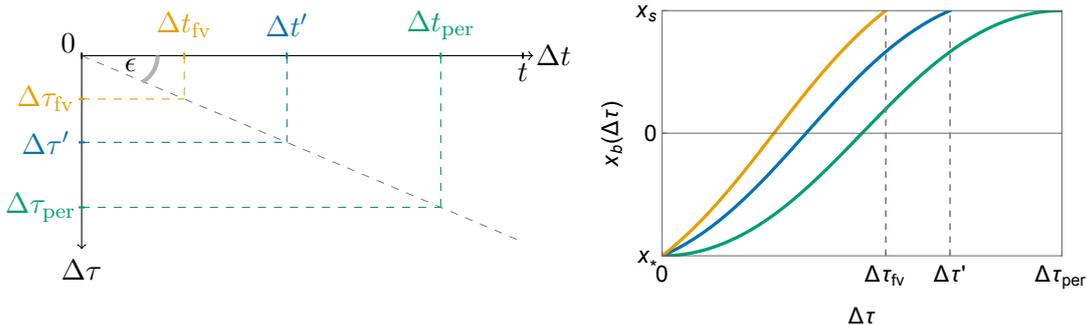
\begin{figure}[t!]
\begin{minipage}[c]{0.52\textwidth}
\centering
\vspace*{-0.5cm}
\begin{tikzpicture}[scale=1.35]
    \draw[->](0,0)--(4.4,0); 
    \draw[->](0,0)--(0,-1.9); 
    \draw[gray, dashed](0,0)--(4.3,-1.8298); 
    \draw[black, thick](0,0.03)--(0,-0.03);
    \draw (-0.13,0.13) node [black]{$0$};
    \draw (4.3,-0.16) node [black]{$t$};
    \draw (4.6,0) node [black]{$\Delta t$};
    \draw[black, thick](4.3,0.03)--(4.3,-0.03);
    \draw (0,-2.1) node [black]{$\Delta \tau$};
    \draw (0.475,-0.09) node [black] {$\epsilon$};
    \draw [gray!60,ultra thick](0.74,0) arc [start angle=0, end angle=-60, radius=0.3];
    \draw[oiOrange, dashed](1,0)--(1,-0.4255); 
    \draw[oiOrange, dashed](0,-0.4255)--(1,-0.4255);
    \draw[oiOrange, thick](1,0.03)--(1,-0.03);
    \draw[oiOrange, thick](-0.03,-0.4255)--(0.03,-0.4255);
    \draw (-0.35,-0.4255) node [text=oiOrange]{$\Delta \tau_{\rm fv}$}; 
    \draw (1,0.3) node [text=oiOrange]{$\Delta t_{\rm fv} $}; 
    \draw[oiBlue, dashed](2,0)--(2,-0.8511); 
    \draw[oiBlue, dashed](0,-0.8511)--(2,-0.8511);
    \draw[oiBlue, thick](2,0.03)--(2,-0.03);
    \draw[oiBlue, thick](-0.03,-0.8511)--(0.03,-0.8511);
    \draw (-0.35,-0.8511) node [text=oiBlue]{$\Delta \tau^\prime$}; 
    \draw (2,0.3) node [text=oiBlue]{$\Delta t^\prime $}; 
    \draw[oiGreen, dashed](3.5,0)--(3.5,-1.4894); 
    \draw[oiGreen, dashed](0,-1.4894)--(3.5,-1.4894);
    \draw[oiGreen, thick](3.5,0.03)--(3.5,-0.03);
    \draw[oiGreen, thick](-0.03,-1.4894)--(0.03,-1.4894);
    \draw (-0.4,-1.4894) node [text=oiGreen]{$\Delta \tau_{\rm per}$}; 
    \draw (3.5,0.3) node [text=oiGreen]{$\Delta t_{\rm per} $}; 
\end{tikzpicture}
\end{minipage}
\begin{minipage}[c]{0.43\textwidth}
\includegraphics[width=\textwidth]{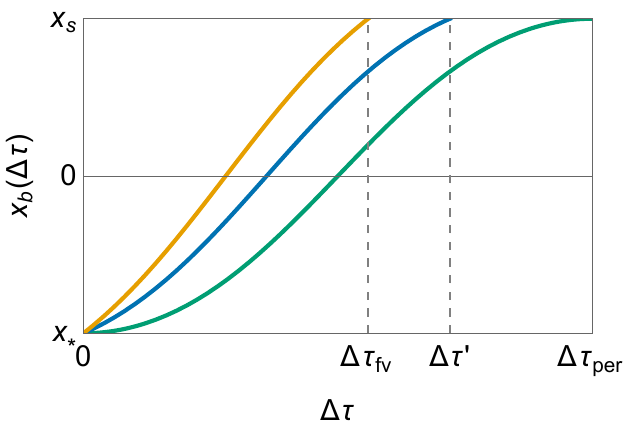}
\end{minipage}
\caption{Eq.~\eqref{eq:P-flux-turning-point} visualised. \textit{Left panel}: For different values of the auxiliary variable $\Delta t$, the saddle points of the path integrals in Eq.~\eqref{eq:P-flux-master} are given by different steadyons. The projection of these solutions onto the imaginary-time axis can be identified with instanton defined on the imaginary-time interval $\Delta \tau \simeq \epsilon \cdot \Delta t$. \textit{Right panel}: Examples for such instantons for the concrete example of the double-well potential used in appendix~\ref{app:steady-review}, assuming a vanishing initial velocity for $x_{\rm Re}$. The instanton corresponding to the shortest depicted $\Delta \tau$, $\Delta \tau_{\rm fv}$, can be identified with a section of the instanton connecting the false vacuum to its corresponding convergence point. While this solution amounts to the smallest Euclidean action, the total integrand in Eq.~\eqref{eq:P-flux-master} is also sensitive to the wave function factor. In Sec.~\ref{sec:resonance}, we demonstrate how this integral can be evaluated for the concrete example of a resonance state.}
\label{fig:dotPR}
\end{figure}

Thus, for any such pair of values $\{\Delta t,\Delta t^\prime \}$, the probability flux $\dot{P}_{\mathcal{R}}$ can be expressed through the actions of its corresponding steadyon pair, $S (\Delta t^{(\prime)})\equiv S_{\rm Re} (\Delta t^{(\prime)})+i S_{\rm Im} (\Delta t^{(\prime)})$, as
\begin{align}
    \dot{P}_{\mathcal{R}} \sim & \int_0^t \d \Delta t\ \psi (x_*,t-\Delta t)   \exp \left[ i S_{\rm Re} (\Delta t)  - S_{\rm Im} (\Delta t)  \right]  \nonumber \\ 
    &\cdot\int_0^t \d \Delta t^\prime \ \psi^* (x_*,t-\Delta t^\prime) \exp \left[-i S_{\rm Re} (\Delta t^\prime) - S_{\rm Im} (\Delta t^\prime) \right] \nonumber \\ 
    =& \left| \int_0^t \d \Delta t\ \psi (x_*,t-\Delta t)   \exp \left[ i S_{\rm Re} (\Delta t)  - S_{\rm Im} (\Delta t)  \right]  \right|^2\,. \label{eq:P-flux-turning-point}
\end{align}
Here, as in the remainder of this article, the symbol $\sim$ indicates that the $\mathcal{O}(1)$ prefactor has been dropped. We finally note that the actions depend on $\Delta t$ and $\Delta t^\prime$, respectively, in two ways: Directly through the length of the time intervals, and indirectly through the change in the solution necessary to accommodate the modified boundary conditions. In the established Euclidean-time picture, this can be understood as a different $\Delta t$ leading to a different $\Delta \tau$, requiring a different initial Euclidean-time momentum of the instanton. See Fig.~\ref{fig:dotPR}.

The form of the remaining expression now suggests to also evaluate the remaining integrals over $\Delta t$ and $\Delta t^\prime$ using a stationary phase approximation once a wave function has been specified. We demonstrate this in Sec.~\ref{sec:resonance} for the example of a resonance state. In Appendix~\ref{app:nonlinear}, we lay out how our analysis based on the direct approach would have to be extended to also incorporate multi-instanton contributions. 

\section{The decay rate of resonance states} \label{sec:resonance}
We are now ready to consider the particularly relevant example of tunneling out of a resonance state, which have been studied extensively in the literature. See, for instance, Refs.~\cite{Kobzarev:1974cp,Brezin:1976vw,Callan:1977pt,Zinn-Justin:2002ecy,Andreassen:2016cvx}. Most important for our present discussion, they are approximate eigenstates of the Hamiltonian satisfying
\begin{align}
    H | \tilde{E}_i \rangle \simeq & \left(E_i - i\frac{\Gamma_i}{2} \right) |\tilde{E}_i \rangle \, .
\end{align}
These are not exact eigenstates of $H$ because they are constructed using a different set of boundary conditions that, effectively, define a different, non-Hermitian Hamiltonian $\tilde{H}$. The latter is defined on $\mathcal{F}$ with outgoing (Gamow-Siegert) boundary conditions at the barrier with $\mathcal{R}$~\cite{Gamow:1928zz,Siegert:1939zz}. They do, however, define a (non-orthogonal) basis of states within $\mathcal{F}$.

As necessary for our discussion, the definition of these states requires that they are localised inside $\mathcal{F}$, and in the semiclassical limit most of their probability is located between two classical turning points with $V(x_*)=E_i$ (beyond which it decays exponentially). The small imaginary part meanwhile is precisely the decay rate we ultimately wish to calculate, 
\begin{align}
    \langle \tilde{E}_i (t) |\tilde{E}_i (t) \rangle \simeq & \left| e^{-i\left(E_i - i\frac{\Gamma_i}{2} \right) t}  |\tilde{E}_i  \rangle \right|^2 \simeq e^{- \Gamma_i t}\langle \tilde{E}_i  |\tilde{E}_i  \rangle \,.
\end{align}
Importantly, as these states are only approximate eigenstates, both of these relations are only valid up to some time $t_{\rm res}$, after which the dynamics (driven by $H$) has propagated enough probability to the region $\mathcal{R}$ to make it impossible to decompose the state only in terms of $\{| \tilde{E}_i \rangle\}$. To be clear, at all times the dynamics is described by unitary time evolution $\sum_i c_i e^{-i \varepsilon_n t} | n \rangle $ where $\{ |n \rangle, \varepsilon_n \}$ are the exact eigenstates and eigenvalues of $H$; however, there is a time window (controlled by how much probability has flowed out of $\mathcal{F}$) where the resonance states describe the dynamics. For the purposes of the following discussion, we shall assume that we have a system for which this time is long enough for this setup to accurately describe the tunneling.\footnote{Note that, as the false vacuum state can be understood as just the lowest-lying resonance state, all of our following statements also apply for the tunneling rate out of this state.} 

On the other hand, as stated in Eq.~\eqref{eq:P-flux-turning-point}, we have an explicit expression that allows us to calculate the tunneling rate $\Gamma_i$ through the probability flux $\dot{P}_{\mathcal{R}}$ as 
\begin{align}
    \Gamma_i = - \frac{\dot{P}_{\mathcal{F}}}{P_{\mathcal{F}}}= \frac{\dot{P}_{\mathcal{R}}}{P_{\mathcal{F}}} \, . \label{eq:gamma-sec3-ratio-of-Pdot-P}
\end{align}
It is then natural to apply this framework to the calculation of the decay rate of a resonance. In fact, since -- by construction -- resonance states have a well-defined decay rate, one can argue that, if anywhere, this is where the derivation that leads to Eq.~\eqref{eq:P-flux-turning-point} should be at the optimum of its applicability. The problem of tunneling out of an arbitrary state is thus reduced to determining whether one can establish a one-to-one correspondence between the calculation of $\Gamma_i$ and a resonance. For notational simplicity, we will in the following drop the index ``$i$'' from the energy and tunneling rate, and focus on establishing the connection for a single resonance state with $\tilde{E} = E - i \Gamma/2$.

A further simplification arises from the tandem of the steadyon picture and the semi-classical nature of our analysis. In the semi-classical picture, tunneling rates are always exponentially smaller than the typical energy scale of the system $\Gamma \sim E \cdot e^{-S_E} $, suppressed so by the value of a Euclidean action obtained via extremisation with respect to the particle path. In order to find these extremal configurations (in fact, to render them well-defined), the steadyon picture introduces a regulator parameter $\epsilon$ that modifies the Hamiltonian $H \to (1-i\epsilon) H$, and tunneling ``steadyon'' solutions can be shown to exist and feature an effective imaginary time evolution of $\Delta \tau = \epsilon \cdot \Delta t$, where $\Delta \tau$ is of the same order of magnitude as the typical timescale of the system (e.g., $\Delta \tau$ may correspond to (half) the time interval of the usual Euclidean ``bounce'' solutions). Thus, $ \epsilon \cdot E \cdot \Delta t = E \cdot \Delta \tau \sim \mathcal{O}(1)$, whereas the calculation in Eq.~\eqref{eq:P-flux-turning-point} requires $\Gamma\cdot \Delta t \ll 1$, or equivalently,
\begin{equation}
    \Gamma \ll \epsilon \cdot  E \, .
\end{equation}
This condition justifies in what follows that we can drop $\Gamma$ from the time evolution altogether, as the imaginary time evolution (with the regulated Hamiltonian) is dominated by the contribution proportional to $\epsilon$, and thus obtain $\Gamma$ as a calculated quantity out of Eq.~\eqref{eq:P-flux-turning-point}. As a side effect of this, the states we are dealing with for practical purposes are nothing more than WKB energy eigenstates of the false vacuum region, which allow for an intuitive grasp of the calculation.\footnote{As such, they are indeed exponentially localised in $\mathcal{F}$ and furnish an approximate basis of states therein.}

\subsection{Resonance states in the steadyon picture}

In light of the preceding discussion, the time evolution of a resonance state in the steadyon picture is nothing more than that implied by the real part of the state's energy, deformed into the complex time plane via $\epsilon$. The net effect is that
\begin{align}
    \psi (t,x)\equiv \langle x| e^{-i H t} |\tilde{E}\rangle \to \langle x| e^{-i H (1 - i \epsilon )t} |\tilde{E}\rangle \simeq e^{-i E t-  \epsilon E t} \psi (x) \, , \label{eq:wavefun-reso-steady}
\end{align}
and, as such, the wave functions in Eq.~\eqref{eq:P-flux-turning-point} then satisfy
\begin{align}
    \psi (x_*,t-\Delta t)  =& e^{-iE(t-\Delta t) - \epsilon E (t-\Delta t )}  \psi(x_*) \, ,  \\
     \psi^* (x_*,t-\Delta t^\prime) =& e^{iE(t-\Delta t^\prime) - \epsilon E (t-\Delta t^\prime )} \psi^*(x_*) \, .
\end{align}
Keeping in mind that the $\epsilon$ deformation also affects the propagators (and therefore defines the extremal configurations on which the action is evaluated), the probability flux~\eqref{eq:P-flux-turning-point} reduces, by means of the steadyon picture, to
\begin{align}
    \dot{P}_{\mathcal{R}} \sim & |\psi(x_*)|^2 e^{-2 \epsilon E t}  \int_0^t \d \Delta t  \ \exp \left[i E \Delta t +i S_{\rm Re} (\Delta t) - S_{\rm Im} (\Delta t)  + \epsilon E \Delta t \right] \nonumber \\
    & \cdot \int_0^t \d \Delta t^\prime \ \exp \left[-i E \Delta t^\prime - i S_{\rm Re} (\Delta t^\prime)  - S_{\rm Im} (\Delta t^\prime)  + \epsilon E  \Delta t^\prime)\right] \nonumber \\ 
    =&  |\psi(x_*)|^2 e^{-2 \epsilon E t} \left| \int_0^t \d \Delta t  \ \exp \left[i E \Delta t +i S_{\rm Re} (\Delta t) - S_{\rm Im} (\Delta t)  + \epsilon E \Delta t \right]  \right|^2 \,. \label{eq:Pdot-res}
\end{align} 
Just as in its previous form, the integrals over $\Delta t$ and $\Delta t^\prime$ arrange into an absolute value. As the exponent is complex, we can analyse and evaluate the integral using either using a stationary phase approximation (extremising the imaginary part) or a saddle point approximation (extremising the real part). To guarantee consistency of our results, we will in the following demonstrate that both the real \textit{and} imaginary part are extremised by $\Delta t= \Delta t_{\rm per}$, where $\Delta t_{\rm per}$ is the real time interval over which the periodic steadyon is defined. For a brief review of the steadyons' properties relevant for this discussion, see appendix~\ref{app:steady-review}. To do so, we recall that the projection of the periodic steadyon onto the real-time axis coincides with the classical oscillatory solution within $\mathcal{F}$ with turning point $x_*$, which we will call $x_{\rm class}(t)$. Its projection onto the Euclidean-time axis meanwhile is identical to the Euclidean-time instanton starting from $x_*$ at rest, which we will denote by $x_I(\tau)$. The real and imaginary parts of the steadyon's complex action meanwhile coincide with the classical and Euclidean action of these solutions, $S_{\rm Re}=S[x_{\rm class}]$ and $S_{\rm Im}=S_E[x_I]$.

\subsection{Periodic steadyons from real-time action}\label{sec:perstead}

We begin by showing that the imaginary part of the exponent in Eq.~\eqref{eq:P-flux-turning-point} is extremised by the periodic steadyon, corresponding to $\Delta t= \Delta t_{\rm per}$. We consider a small negative perturbation of $\Delta t$ around the expected extremal value, $\Delta t = \Delta t_{\rm per}-\delta t$. To understand the effect of this perturbation on the action, we recall that for $\Delta t=\Delta t_{\rm per}$ the real-time action $S_{\rm Re}$ is that of the classically oscillatory motion with turning point $x_*$, $x_{\rm per}(t)$. The change in the boundary condition would therefore lead to a change in the steadyon, with the projection of the new solution onto the real-time axis taking (to leading order) the form
\begin{align}
    x_{\rm class}(t)\to x_{\rm class}(t) + \delta x(t) \,.
\end{align}
The perturbed action $S_{\rm Re}$ can therefore be written in terms of the unperturbed action $S_{{\rm Re}, {\rm per}}$ as 
\begin{align}
    S_{\rm Re}= & \int_0^{\Delta t} \frac{m}{2}\dot{x}^2 -V(x)\ \d t=\int_0^{\Delta t_{\rm per}-\delta t} \frac{m}{2}\left(\dot{x}_{\rm class} + \delta \dot{x} \right)^2 -V(x_{\rm class}+ \delta x)\ \d t\nonumber \\ 
    \simeq & S_{{\rm Re},{\rm class}} - \int_{\Delta t_{\rm class}-\delta t}^{\Delta t_{\rm class}} \frac{m}{2} \dot{x}_{\rm class}^2 -V(x_{\rm class}) \ \d t + \int_0^{\Delta t_{\rm class}}m\ \dot{x}_{\rm class}  \delta \dot{x} -V^\prime (x_{\rm class})\delta x\ \d t \,.
\end{align}
We can now integrate by parts the last term, which vanishes due to the equation of motion of $x_{\rm per}$. Moreover, this allows us to expand the second time as
\begin{align}\label{eq:Eimp}
    - \int_{\Delta t_{\rm per}-\delta t}^{\Delta t_{\rm per}} \frac{m}{2} \dot{x}_{\rm class}^2 -V(x_{\rm class})\ \d t\simeq  \delta t \cdot V(x_{\rm class}(\Delta t_{\rm per}))= \delta t \cdot V(x_s)= \delta t \cdot E\,.
\end{align}
Meanwhile, we find for the transformation of the wave function phase factor
\begin{align}
    E \cdot \Delta t \to E \cdot  \Delta t_{\rm per} - E \cdot \delta t \, .
\end{align}
Thus, the overall phase transforms as
\begin{align}
    E \cdot  \Delta t +S_{\rm Re}(\Delta t) \to E \cdot  \Delta t_{\rm per} - E \cdot  \delta t + S_{\rm Re}(\Delta t_{\rm per}) + E\cdot  \delta t =E \Delta t +S_{\rm Re}(\Delta t) \, .
\end{align}
This establishes that the integral over $\Delta t$ is indeed dominated by the contribution from $\Delta t=\Delta t_{\rm per}$, as we had initially conjectured. Hence, the Euclidean action reduces to that of the periodic instanton defined on the Euclidean time interval $\Delta \tau_{\rm per}= \epsilon \Delta t_{\rm per}$. Squaring the total expression then leads to an additional factor of $2$, ultimately allowing us to recover the Euclidean action of the periodic bounce $x_{b,\rm{per}}$. Altogether, this implies that the probability flux $\dot{P}_{\mathcal{R}}$ is to leading order given by
\begin{align}\label{eq:PRdotstat}
    \dot{P}_{\mathcal{R}} \sim & e^{-2 \epsilon E t} \exp \left( - S_E [x_{b,\rm{per}}]  + 2  E \Delta \tau_{\rm per} \right) \, .
\end{align}
This indeed agrees with the WKB result, up to the prefactor $e^{-2 \epsilon E t}$. We will show in Sec.~\ref{sec:denom} that this factor is precisely canceled by the denominator on the r.h.s. of Eq.~\eqref{eq:gamma-sec3-ratio-of-Pdot-P}, thus verifying that $\Gamma$ is given by the WKB result.

In principle, we could have arrived at this result through physical reasoning alone. Recall that we introduced the parameter $\epsilon$ as a small deformation of the theory, and the significant deviation from classical behavior can directly be traced back to the limit $t \gg \epsilon^{-1} t_{\rm sys}$. Conversely, for $t \lesssim t_{\rm sys}$, it is reasonable to expect that the preferred saddle point should converge against the classical solution.

Our derivation also makes evident the significance of choosing $x_*$ as the classical turning point: A value closer to the false vacuum would have rendered the approximation that the wave function be localised beyond $x_*$ less accurate, leading to a less reliable tunneling rate. While choosing a value closer to the barrier top would indeed improve this approximation, it would also affect the evaluation of the $\Delta t$-integral. This can be seen clearly in Eq.~\eqref{eq:Eimp}, which depends explicitly on the value of the potential at $x_s$. The condition that $V(x_s)=V(x_*)$ is then enforced by the properties of the periodic steadyon.

\subsection{Periodic steadyons from Euclidean-time action}

A similar argument applies to the real part of the exponent, establishing that the integrals over $\Delta t$ and $\Delta t^\prime$ can equivalently be evaluated through a saddle point approximation rather than a stationary phase approximation. For the same reason as the periodic steadyon's projection on the real-time axis was a solution to the real-time equations of motion, its projection onto the Euclidean-time axis solves the Euclidean equations of motion. Moreover, the small variation in the integration variable $\Delta t$ translates to a small variation in the Euclidean time interval on which this solution is defined, $\Delta \tau = \epsilon \cdot (\Delta t_{\rm per} - \delta t)= \Delta \tau_{\rm per} - \delta \tau$. Thus, the perturbation affects the imaginary part of the action as
\begin{align}
    S_{\rm Im}=& \int_0^{\Delta \tau} \frac{m}{2}\dot{x}^2 + V(x) \d \tau = \int_0^{\Delta \tau_{\rm per} - \delta \tau} \frac{m}{2}(\dot{x}_I+\delta\dot{x})^2 + V(x_I + \delta x) \d \tau \nonumber \\ 
    \simeq & S_{{\rm Im},{\rm per}} - \int_{\Delta \tau_{\rm per} - \delta \tau}^{\Delta \tau_{\rm per}} \frac{m}{2}\dot{x}_I^2 + V(x_I )\ \d \tau + \int_0^{\Delta \tau_{\rm per} } m \dot{x}_I  \delta\dot{x} + V^\prime(x_I )\delta x \ \d \tau \, .
\end{align}
As for the real-time action, the last term vanishes due to the Euclidean equations of motion and the boundary conditions of the instanton. For the second term, meanwhile, we have
\begin{align}
    - \int_{\Delta \tau_{\rm per} - \delta \tau}^{\Delta \tau_{\rm per}} \frac{m}{2}\dot{x}_I^2 + V(x_I ) \d \tau\simeq - \delta \tau \cdot E \, .
\end{align}
The contribution from the wave function transforms trivially as $E \cdot \Delta \tau \to E \cdot \Delta \tau - E \cdot  \delta \tau$, so that we find for the real part of the exponent 
\begin{align}
    -S_{\rm Im}+ E \cdot\Delta \tau \to - (S_{\rm Im} -\delta \tau \cdot E )+ E \cdot \Delta \tau - E \delta \tau = -S_{\rm Im}+ E \cdot\Delta \tau \, .
\end{align}
Thus, the periodic steadyon indeed extremises the both the real and imaginary part of the exponent with respect to $\Delta t$, confirming Eq.~\eqref{eq:PRdotstat}.

\subsection{Denominator} \label{sec:denom}

Having evaluated $\dot{P}_{\mathcal{R}}$ using the steadyon picture, it only remains to calculate $P_{\mathcal{F}} (t)$ in the deformed theory in order to determine $\Gamma$ using Eq.~\eqref{eq:gamma-sec3-ratio-of-Pdot-P}. For a resonance state in the semiclassical approximation, this is straightforward using Eq.~\eqref{eq:wavefun-reso-steady}:
\begin{align}
    P_\mathcal{F} (t)= \int_\mathcal{F} |\psi (t,x)|^2 \d x= \int_{\mathcal{F}} e^{- 2 \epsilon E t} |\psi (x)|^2 \simeq e^{- 2 \epsilon E_ t} \, .
\end{align}
Note that the condition $\Gamma \ll \epsilon E$ allowed us to omit $\Gamma$ in the time evolution factors without adverse consequences (conversely, if we went outside the regime $\Gamma \Delta t \ll 1$, the approximations we used to do the calculation would not be warranted). Remarkably, the additional factor cancel precisely with the $t$-dependent term in $\dot{P}_{\mathcal{R}}$, leading us to recover the established result obtained using the WKB method:
\begin{align}\label{eq:WKB}
    \Gamma  =  \frac{\dot{P}_{\mathcal{R}}(t)}{P_{\mathcal{F}}(t)} \simeq  |\psi(x_*)|^2 \exp \left( - S_E [x_{b,\rm{per}}]  + 2  E \Delta \tau_{\rm per} \right) \, .
\end{align}

\subsection{Section summary}

In summary, we have shown that it is possible to calculate the decay rate of (approximate) energy eigenstates using the steadyon picture, reproducing the well-known WKB results. In the semiclassical approximation, there is a priori no restriction to the value of the energy $E$ and the decay rate $\Gamma$ may be obtained as a correction via the calculation we outlined above, analogously to how a decay rate can be calculated in the WKB approximation as a function of the energy. This picture has the advantage that the calculation is direct (in the same sense as in~\cite{Andreassen:2016cff,Andreassen:2016cvx}), and the outgoing boundary conditions for the wavefunction are explicitly enforced by requiring that the paths in the functional integral cross the turning points without going back.

The nontrivial connection we made in this section is to show that the extremal path configurations coming out of the steadyon analysis when coupled to an (approximate) energy eigenstate of the false vacuum basin give the same result as the WKB analysis, even though the theory has been deformed and therefore the paths are fully complex. Viewed from the other end, we have identified how it is that ``natural-looking'' path configurations -- concretely, the periodic steadyon -- gets selected from the path integral by the boundary conditions imposed by the energy eigenstate. 

Overall, this provides nontrivial insight into tunneling processes in real time. To further substantiate our discussion, and also test the validity of the semiclassical approximation in practice, we now turn to numerical studies of a particle tunneling out of a potential in real time.

\section{Numeric Simulations} \label{sec:numbers}

The analysis in Secs.~\ref{sec:tunnel-steady} and~\ref{sec:resonance} suggests that the decay rate of arbitrary states localised within a false vacuum basin $\mathcal{F}$ can be understood by decomposing the state into resonant states, where each resonance decays with an exponential rate approximately given by the corresponding WKB rate. In this section, we test this conclusion by comparing resonant expansion fits to numerical simulations of decays of initial states localised within $\mathcal{F}$. The assumption used in this section is that resonant states $|\tilde{E}_i \rangle$ form a basis for generic states $|\Psi\rangle$ localised within $\mathcal{F}$, such that it is possible to decompose $|\Psi\rangle$ as
\begin{align}
    |\Psi\rangle= \sum_{i=1}^\infty \psi_i | \tilde{E}_i \rangle \, ,
\end{align}
where their time evolution of each resonant state $|\tilde{E}_i\rangle$ follows by multiplying the resonance by the non-unitary time evolution factor $e^{-i\tilde{E}_i t }= e^{- i E_i t - \Gamma_i t/2}$.

A distinctive feature of this decomposition is that the basis states $|\tilde{E}_i \rangle$ are, in general, not orthogonal, as the outgoing boundary conditions used to define resonance states break the hermiticity property of the Hamiltonian $H$. If the basis states were orthogonal, one would expect the time evolution of the probability of finding the state in $\mathcal{F}$ as a function of time to be given by a sum of decaying exponentials,
\begin{equation}\label{eq:spectraldecomp}
P_\mathcal{F}(t) = \sum_{i=1}^\infty |\psi_i|^2 e^{-\Gamma_i t} \, .
\end{equation}
Eq.~\eqref{eq:spectraldecomp} is essentially identical to a spectral decomposition of a two-point correlation function for a Euclidean time Quantum Field Theory (QFT) (for example, Lattice-QCD~\cite{Gattringer:2010zz,Montvay:1994cy}), and fitting the overlap factors $\psi_i$ and decay constants $\Gamma_i$ given $P_\mathcal{F}(t)$ could be performed in a similar fashion~\cite{Lin:2007iq,Wagman:2024rid,Fox:1981xz}. 
Instead, due to the non-orthogonality of the resonant states, the probability of finding the particle in the false vacuum is given by
\begin{align}\label{eq:discretePF}
    P_{\mathcal{F}}(t) &= \sum_{ij} \psi_j^* \langle \tilde{E}_j | \tilde{E}_i \rangle \psi_i \,  e^{-i(E_i - E_j) t} e^{-(\Gamma_i + \Gamma_j)t/2} \, .
\end{align}
where the non-orthogonality of the resonant states $|\tilde{E}_i \rangle$ introduces oscillations into the the observed $P_\mathcal{F}(t)$. 
Given such a resonant expansion, it is possible to fit an observed decay rate $P_\mathcal{F}(t)$ to extract the complex numbers $\psi_i, \langle \tilde{E}_j | \tilde{E}_i \rangle$, and the real parameters $E_i, \Gamma_i$. In the absence of back-tunneling (which can be arranged in numerical simulations by preparing a suitably wide region $\mathcal{R}$), the decay constants are all positive ($\Gamma_i > 0$) causing the large-time behavior of $P_\mathcal{F}(t)$ to be dominated by the resonant states with the smallest decay constants. 

\begin{figure}[t]
    \centering
    \includegraphics[width=1.0\linewidth]{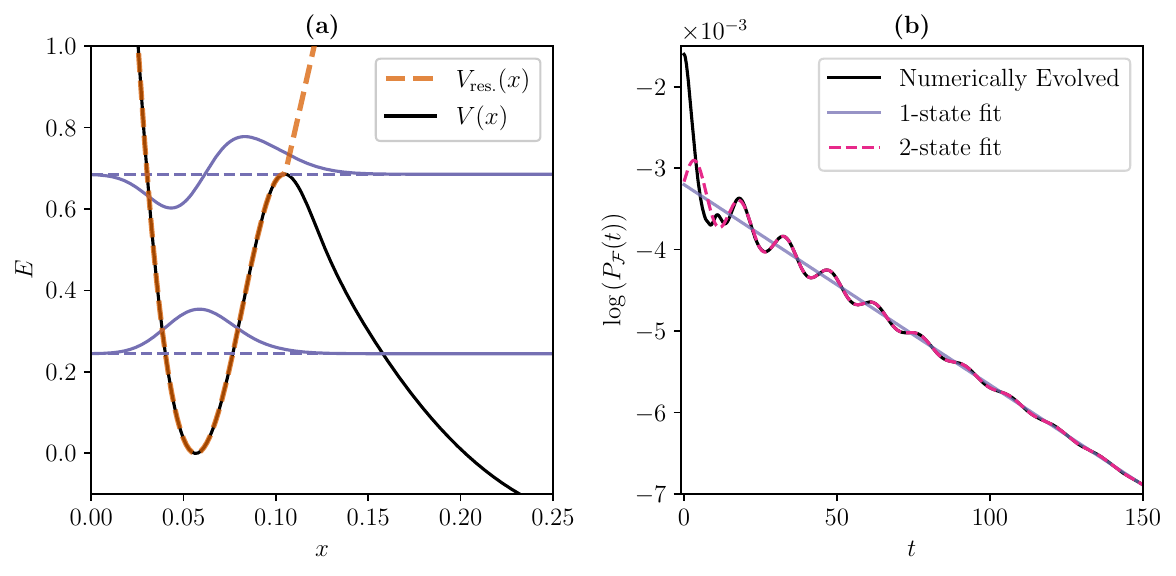}
    \caption{\textbf{(a)} Potential used in these numeric simulations is shown as $V(x)$. The potential $V_\mathrm{res}(x)$ defined in \Cref{eq:Vres} is also shown. The first four resonant states with $m = 6\cdot10^3$ (in natural units $\hbar = 1$), displaced vertically by their respective energies shown by the dashed lines are plotted in the figure. \\
    \textbf{(b)} The lowest approximate resonant state was time evolved over a total time $T = 150$, and the overlap onto the false-vacuum is plotted in black. Fits to \Cref{eq:discretePF} are shown both for only including a single state, as well as including two states in the fit. }
    \label{fig:j1}
\end{figure}

To provide numerical evidence of this decomposition, we carry out simulations of resonance tunneling for a potential shown in Fig.~\ref{fig:j1}(a), which was inspired by the potential used in Refs.~\cite{Andreassen:2016cvx,Andreassen:2016cff}\footnote{Refs.~\cite{Andreassen:2016cvx,Andreassen:2016cff} do not provide an explicit form of the potential used. Our potential is given by digitising their potential and fitting with cubic splines, along with a modification to enlarge the true vacuum basin. Explicitly, $V(x)$ is given by a cubic spline interpolation of the following $(x,V(x))$ tuples : $\{(0.03, 0.7),
 (0.05, 0.035),
 (0.075, 0.21),
 (0.1, 0.67),
 (0.125, 0.52),
 (0.15, 0.3),
 (0.2, 0.01),
 (0.25, -0.14),$
$ (0.3, -0.19),
 (5.4, -0.21),
 (5.6, -0.15),
 (5.8, 0.055),
 (5.9, 0.295),
 (5.95, 0.5),
 (6.0, 1.0)\}.$ } except with a larger true vacuum basin $\mathcal{R}$ to allow for longer periods of time for the solution to decay without scattering off the right-hand side of the true vacuum basin.
We take the false-vacuum region $\mathcal{F}$ in this section to be the region with $x \leq x_m$, where $x_m$ is the location of the local maxima of the potential $V(x)$. Note that this differs from the definition used in previous sections, however the difference is given by the norm of the wavefunction at the barrier (which is exponentially suppressed in the barrier height) and largely does not alter the observed decay rates. For such a potential that doesn't have a simple analytic form (unlike the square potential well considered in Refs.~\cite{Andreassen:2016cvx,Andreassen:2016cff} and the quartic potential as considered in Ref.~\cite{Steingasser:2024ikl}), it is not feasible to exactly calculate the resonant states and their associated decay constants. It is possible to \textit{approximately} describe the resonant states as eigenstates of the Hamiltonian associated to a modified potential, where the true vacuum has been replaced by a confining potential that causes the eigenstates to be localised within $\mathcal{F}$. For the states shown in Fig.~\ref{fig:j1}(a), we used the modified potential
\begin{equation}\label{eq:Vres}
V_\mathrm{res}(x) = \begin{cases}
    V(x) &  \mathrm{for} \quad x \leq x_0 \\
    V(x_0) + \lambda (x-x_0)  & \mathrm{for} \quad x > x_0
\end{cases} \, ,
\end{equation}
with $\lambda = 20$. The approximate resonant states are shown in blue. Note that states higher up in the spectrum are likely not to be trustworthy, as they have energies larger than the potential barrier. Instead, physically one should expect a continuum of scattering states traveling to the right with energies above the potential barrier. 

In Fig.~\ref{fig:j1}(b), we present the lowest energy approximate resonance's overlap onto $\mathcal{F}$ as a function of time, where the false-vacuum region $\mathcal{F}$ is defined by the region to the left of the potential barrier's local maximum. We perform the numeric evolution of the Schrödinger equation by exact diagonalisation of a lattice-discretised Hamiltonian, where the lattice spacing is given by $\mathrm{d}x = 0.0016$. As a cross-check, we also compared these results against an explicit fifth-order Runge-Kutta~\cite{DORMAND198019} discretisation of the Schrödinger equation, discretising the $1$-dimensional Laplacian with a 7-point stencil, finding no substantial difference to exact diagonalisation. To demonstrate the utility of the resonant decomposition~\eqref{eq:discretePF}, we fit the ansatz $\log(P_\mathcal{F}(t))$ with either one or two resonant states included in the fit. Note that including only one state in the resonant decomposition leads to the form
\begin{equation}
\log\left[ P_\mathcal{F}(t)\right] = \log \left[ |\psi_1|^2\langle \tilde{E}_1 | \tilde{E}_1 \rangle \right] - \Gamma_1 t \, ,
\end{equation}
which is an exact exponential decay as a function of time. If the approximate resonant states were exact, there would be no `contamination' from higher-energy resonant states, and the observed $\log (P_\mathcal{F}(t))$ would be linear in $t$. 
Instead, the one-state fit shown in Fig.~\ref{fig:j1}(b) becomes more accurate at larger times as all the higher-energy resonant state contributions have sufficiently decayed. At intermediate times in Fig.~\ref{fig:j1}(b) one can see oscillations in $P_\mathcal{F}(t)$, which can be faithfully described by including a second  state in the resonant expansion \Cref{eq:discretePF}. 

\begin{figure}[t]
\centering
    \includegraphics[width=1.0\linewidth]{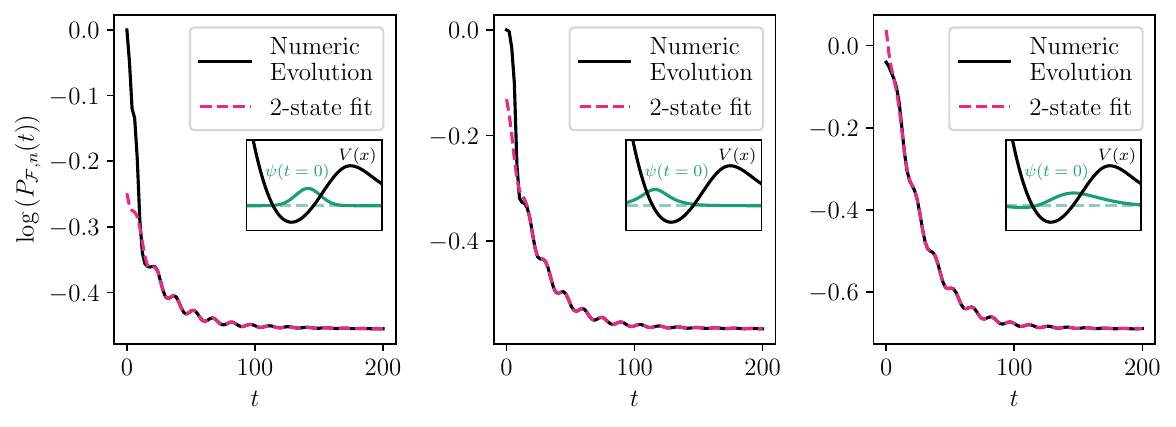}
    \caption{Simultaneous fit to the time evolution of various different intial states. The inset panels on each subfigure show the initial state used in each simulation plotted on top of the potential $V(x)$. The data for all times $t \geq 60$ serves as input for the simultaneous fit, for which the energies $E_i$ and decay constants $\Gamma_i$ were held constant between the different simulations. As the three different initial states had differing amounts of contamination from (unknown) higher excited resonances, each simulation has a different time beyond which the excited resonances have decayed enough for the two-state fit to be a good description.}
\label{fig:simfit}
\end{figure}

To demonstrate the robustness of these fits, we perform a simultaneous fit to the decay $P_{\mathcal{F},n}$ of \textit{multiple} different initial states, indexed by the label $n = 1,\dots,n_\mathrm{in}$, where $n_\mathrm{in}$ is the number of different initial states used, and $n_\mathrm{s}$ is the number of resonant states included in the fit\footnote{This is a generalisation of the Generalised Eigenvalue Method commonly used in analyses of Lattice-QCD correlation functions to a situation where the energies are allowed to be complex and the basis of states is not orthogonal~\cite{Blossier:2009kd}.}. 
Explicitly, the fit-form is given by
\begin{equation}\label{eq:fitform}
P_{\mathcal{F},n}= \sum_{i,j=1}^{n_\mathrm{s}} {\rm Re} \left\{ \psi_{j,n}^* \langle \tilde{E}_j | \tilde{E}_i \rangle \psi_{i,n} \, e^{-i(E_i - E_j) t} \right\} e^{-(\Gamma_i + \Gamma_j)t/2} \, .
\end{equation}
Counting the number of \textit{real} parameters in such a fit, it would a priori appear that there are $2 n_\mathrm{in} n_s$ parameters required for the occupation numbers $\psi_{i,n}$, $n_s^2$ parameters for the hermitian matrix $\langle \tilde{E}_j | \tilde{E}_i \rangle$, $n_s$ parameters for the real energies $E_i$ and $n_s$ parameters for the real decay constants $\Gamma_i$. However, there are certain redundancies in this description, encapsulated in the following three transformations:
\begin{align}\label{eq:redundant}
\circled{1} &: |\tilde{E}_i \rangle \mapsto \psi_{i,1} |\tilde{E}_i \rangle, \quad \psi_{i,1} \mapsto 1, \qquad   
\circled{2} : \psi_{i,n} \mapsto e^{i \theta_n} \psi_{i,n} , \qquad \circled{3} : E_i \mapsto E_i + c
\end{align}
Under any of the transformations described in Eq.~\eqref{eq:redundant}, $P_{\mathcal{F},n}$ is invariant for all $n$.\footnote{There are various other ways to express the same transformations, for example instead of transformation {\tiny $\circled{1}$} that absorbs one set of occupation numbers into the $\langle \tilde{E}_i | \tilde{E}_j \rangle$ matrix, one can instead set the diagonal of the $\langle \tilde{E}_i | \tilde{E}_j \rangle$ matrix to $1$ (normalising the resonant states) and then using local phase rotations to set $\psi_{i,1}$ to all be real numbers.} The total number of remaining, non-redundant real parameters is thus given by $(n_\mathrm{in}-1) (2n_s - 1) + n_s^2 + 2n_s - 1.$ Furthermore, there is a discrete transformation given by
\begin{equation}
\circled{4} : E_i \mapsto -E_i, \quad \psi_{i,n} \mapsto \psi_{i,n}^*, \quad \langle \tilde{E}_j | \tilde{E}_i \rangle \mapsto \langle \tilde{E}_j | \tilde{E}_i \rangle^* \,.
\end{equation}
To break this discrete degeneracy, in all fits to two states shown in this section it we assume that $\Gamma_1 < \Gamma_2$ implies $E_1 < E_2$, as lower energy resonances are expected to tunnel slower through the barrier. Transformation $\circled{3}$ implies that only energy differences can be extracted from fits to $P_{\mathcal{F},n}$. Decay constants $\Gamma_i$ are invariant under all redundancies, as well as certain combinations of parameters such as the real overlap constants $O_{ij}$ and the real contributions $\phi_i$,
\begin{equation}
O_{ij} := \sqrt{\frac{\langle \tilde{E}_j | \tilde{E}_i \rangle \langle \tilde{E}_i | \tilde{E}_j \rangle}{\langle \tilde{E}_j | \tilde{E}_j \rangle \langle \tilde{E}_i | \tilde{E}_i \rangle } }, \quad \phi_{i,n} := \sqrt{\psi_{i,n}^* \langle \tilde{E}_i | \tilde{E}_i \rangle \psi_{i,n}}\,.
\end{equation}
In Fig.~\ref{fig:simfit}, we present fits to the three different initial states shown in green in the inset panels. Clearly, two resonant states are sufficient to explain the long-time oscillations and decay of $P_{\mathcal{F},n}(t)$. For the fit shown in Fig.~\ref{fig:simfit}, the fitted parameters are given by
\begin{equation}\label{eq:ext1}
\Gamma_1 = 2.414 \cdot 10^{-5}, \quad \Gamma_2 = 4.845 \cdot 10^{-2}, \quad E_2 - E_1 = 0.427, \quad 
    O_{12} = 0.022 \,.
\end{equation}
We estimate that the error bars on these quantities are negligible, though in principle a systematic error (due to contributions from higher energy resonant states that are not included in the fit form) could be estimated by varying over different fit ranges. Note that the limit that $\Gamma \to 0$ is the limit in which the potential barrier extends to $+ \infty$, and the resonant states become true eigenstates that are orthogonal. 
Thus, one should expect that the off-diagonal overlap factors $O_{ij}$ are suppressed in the $\Gamma \to 0$ limit, which is confirmed by the small observed value for $O_{12}$ in the simultaneous fit. 

\begin{figure}[t]
    \centering
    \includegraphics[width=1.0\linewidth]{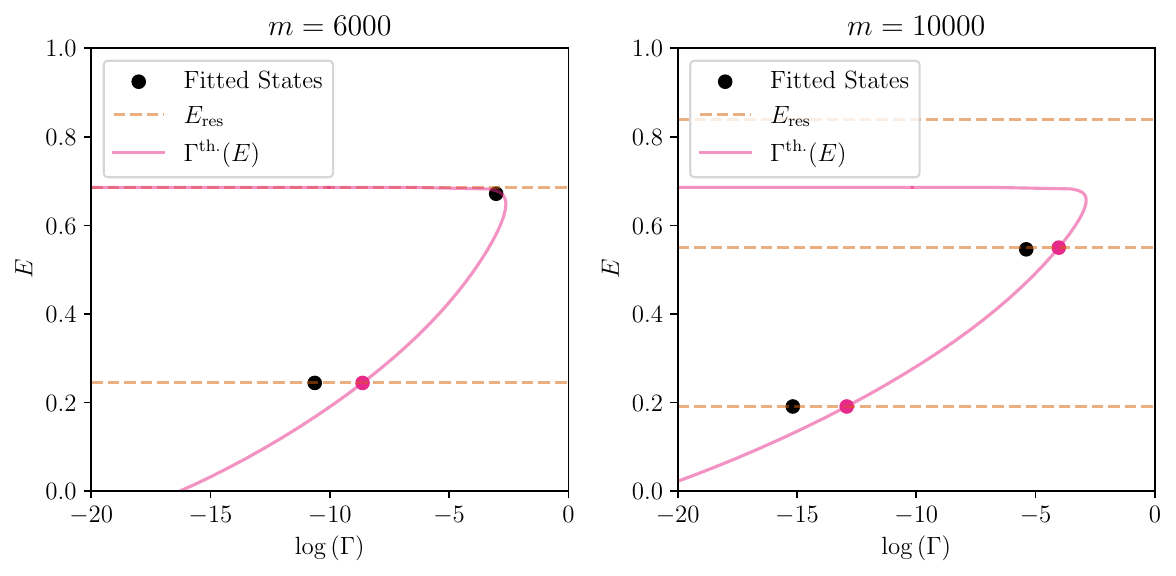}
    \caption{Comparison between the theory prediction provided in \Cref{eq:thadhoc} to the numerically extracted states in the $(E,\log(\Gamma))$ plane for two different masses. $E_\mathrm{res}$ are the energies given be exact diagonalisation of the modified potential~\eqref{eq:Vres}. As only energy shifts can be extracted by fits to $P_{\mathcal{F},n}$, all energies have been shifted such that the lowest energy fitted state has the energy given by the lowest energy $E_\mathrm{res}$. }
    \label{fig:WKBcomp}
\end{figure}

Note that the WKB-like  expressions derived in Eq.~\ref{eq:WKB} for the decay rates of resonances do not include a derivation of an associated dimensionful `prefactor', required for the theoretical rate $\Gamma^\mathrm{th.}$ to be dimensionfully correct. In the absence of this derivation, we modify the dimensionful prefactor derived in Ref~\cite{Callan:1977pt,Zinn-Justin:2002ecy} in an ad-hoc way in the following refinement of Eq.~\ref{eq:WKB} in order to perform comparisons against the numerical data,
\begin{equation}\label{eq:thadhoc}
\Gamma^\mathrm{th.} (E) = \left( \frac{[\frac{\partial^2 V}{\partial x^2} ]^3 (x_s-x_*)^4}{16 m \pi^2}\right)^\frac{1}{4}\mathrm{exp} \left( -{2}\int_{x_-}^{x_+} \sqrt{2m (V(x) - E)}\right) \,,
\end{equation}
where $V(x_*) = V(x_s) = E$ are the classical turning points in the inverted potential, with $x_*$ in $\mathcal{F}$ and $x_s$ on the other side of the barrier (such that $x_* < x_s$ for the potential used in this section). The second derivative of $V(x)$ is evaluated in the false vacuum. As the energy $E$ increases, the WKB-like exponential term suppresses $\Gamma^\mathrm{th.}$. However, note that as $E$ approaches the local maximum of the potential barrier, the WKB-like exponential approaches $1$, whereas the prefactor approaches zero linearly, causing $\Gamma^\mathrm{th.}(E)$ to approach zero. 
This behavior causes the theory prediction plotted in \Cref{fig:WKBcomp} to be non-monotonic, but this is expected due to the breakdown of any semi-classical analysis as energies approach the barrier height. 

At $m = 6\cdot10^3$, the second approximate resonant energy $E_\mathrm{res}^{(2)}$ is very close to the barrier height, causing predictive power to break down. However, the first resonant energy is below the barrier, and we find for its corresponding state
\begin{equation}
\Gamma_1^\mathrm{th.} = 1.79 \cdot 10^{-4}, \quad E^{\mathrm{res}}_2 - E^{\mathrm{res}}_1 = 0.441\,.
\end{equation}
This amounts to a good agreement with the extracted parameters given in Eq.~\ref{eq:ext1}, up to an $O(1)$ factor for the decay constant. As the mass is increased more resonant states are confined the barrier height. Taking for example $m = 10^4$ we find two states below the barrier height, with fitted parameters
\begin{equation}
\Gamma_1 = 2.567\cdot10^{-7} , \quad \Gamma_2 = 4.548\cdot 10^{-3}, \quad E_2 - E_1 = 0.355\,.
\end{equation}
The theoretical predictions meanwhile are given by
\begin{equation}
\Gamma_1^\mathrm{th.} = 2.456\cdot 10^{-6}, \quad \Gamma_2^\mathrm{th.}=1.780\cdot10^{-2},\quad E^{\mathrm{res}}_2 - E^{\mathrm{res}}_1 = 0.359\,,
\end{equation}
which display reasonable agreement. Again, it should be stressed that the prefactor in Eq.~\ref{eq:thadhoc} is an ad-hoc prefactor to obtain a dimensionfully correct theoretical prediction for the decay rate. 
A full analysis should also incorporate a prefactor derived using steadyon methods, which has not been attempted in this study. 

\begin{figure}[t]
\centering
    \includegraphics[width=0.7\linewidth]{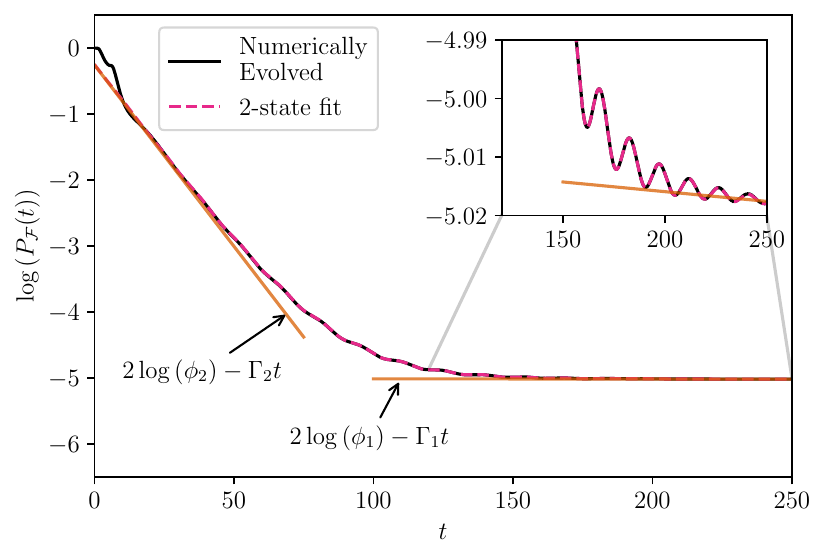}
    \caption{A simulation demonstrating the multi-slope behaviour of $P_\mathcal{F}(t)$, with the same potential, but a different mass $m = 5778.125$, and with an initial state crafted to have large overlap onto the second resonant state. After fitting the numerically evolved $P_\mathcal{F}(t)$ to the 2-state resonant expansion \Cref{eq:fitform} (shown in pink), the individual contributions due to each resonant state can be extracted as $\log(P_\mathcal{F}(t)) = 2 \log(\phi_i) - \Gamma_i t$, and these are plotted as straight orange lines in the figure. The inset plot shows a zoomed in view of the late-time evolution, as the numerical evolution approaches the single-state decay rate given by $\Gamma_1$.}
\label{fig:simfit2}
\end{figure}

To emphasise the fact that decays out of a false-vacuum are characterised by \textit{multiple} resonances (not just a single decay rate), a state can be prepared which predominantly overlaps with the second lowest energy resonant state, which some small overlap onto the lowest energy resonant state. 
The numeric evolution of such a state is shown in Fig.~\ref{fig:simfit2}, which initially decays predominantly with a large slope corresponding to the second decay rate $\Gamma_2$. After the second resonant state has decayed sufficiently, there is an intermediate time window with oscillations in $\log P_\mathcal{F}(t)$ corresponding to mixings between the two resonant states due to their non-orthogonality, followed by a long-time decay rate given by the decay rate $\Gamma_1$ of the lowest energy resonance state. 

\section{Conclusion}

In this work, we have further developed the steadyon picture for tunneling out of a false vacuum, leading to a master formula for the probability flux out of any initial wave function. To demonstrate its application, we have considered approximate (complex) energy eigenstates localised in the false vacuum basin, so-called \textit{resonance states}. We did so via a real-time analysis of the quantum-mechanical path integral, for which we introduced a regulator $\epsilon$ (not very different from the familiar $i\epsilon$ that defines the time-ordered prescription for Feynman propagators in QFT) in order to render the calculations well-defined. We verified that physical rates do not depend on this regulator in the limits of a small $\epsilon$ and physical times large relative to the system's typical time scale, $t \gg t_{\rm sys}$. Overall, this provides a framework for calculations out of excited states in a false vacuum basin directly in terms of a real-time path integral, from which the tunneling rate out of each excited (resonance) state can be calculated. As one might have expected, these rates are in fact equal to the rates one could have calculated from a WKB analysis, for which we now have a path-integral description in real time in terms of the steadyon picture. 

Our analysis also inherits the straightforward and conceptually clear generalisation to higher dimensions intrinsic to the direct approach. In addition to extending the spatial integrals throughout our analysis, this generalisation would only require to replace the points $x_*$ and $x_s$ in Eqs.~\eqref{eq:Dbar},~\eqref{eq:Dunderbar} and~\eqref{eq:Dbarunderbar} with the hypersurfaces of energetically degenerate points on their respective sides of the potential barrier.

On the other hand, we explored the practical feasibility of observing these multiple tunneling rates in solutions to Schr\"odinger's equation for a suitable potential. We demonstrated, for a variety of initial conditions, that the tunneling rates out of the ground and first excited resonance states could be unambiguously identified in the numerical solutions. We also compared these numerically extracted rates with the theoretical prediction from the steadyon analysis, finding reasonable agreement between them. 

We have also addressed a central conceptual question: in principle, one might expect the tunneling out of a \textit{generic initial state} to be dominated by a phase of initial sloshing, followed by a transition to a regime in which the exponential loss of probability is dominated by the contribution from false vacuum state. By finding states whose tunneling behavior shows two clearly identifiable slopes (corresponding to the false vacuum contribution as well as that of the first excited resonance state), we have shown explicitly that this behavior is in fact not universal. Instead, our results establish that it is, in fact, possible to achieve theoretical control over the loss of probability in the regime \textit{before} the contribution of the false vacuum becomes dominant. Our results furthermore show that this can amount to a dominant effect.

Our analysis not only expands the range of applicability of the steadyon picture, but also addresses major previously unanswered questions about its interpretation. It illustrates, in particular, the independence of the final result from the chosen regularisation parameter, and demonstrates how to relate the initial state to the semi-classical steadyon controlling the tunneling rate. In a future work, we will further explore the latter point by considering different examples, for which no analytical results akin to the WKB analysis of the resonance states exists. On a conceptual level, natural questions for future work include the calculation of the path integral prefactors in the steadyon picture as well as its explicit application to quantum fields. 

\section*{Acknowledgements}

We acknowledge the hospitality of the Munger residence and the Kavli Institute for Theoretical Physics at the University of California, Santa Barbara, where parts of this research were conducted. We furthermore thank Matt D. Charles for significant contributions to an inspiring work environment, Sebastian Schenk for helpful discussions and Felix Yu for useful comments on an earlier version of this draft.

Portions of this work were conducted in MIT's \textit{Center for Theoretical Physics - a Leinweber institute} and partially supported by the U.S.~Department of Energy under Contract No.~DE-SC0012567. 
While at MIT, JL's contributions were supported by the U.S. Department of Energy under Contract No.~DE-SC0011090, by the SciDAC5 award DE-SC0023116, and additionally by the National Science Foundation under Cooperative Agreement PHY-2019786 (The
NSF AI Institute for Artificial Intelligence and Fundamental Interactions, http://iaifi.org/). 
While at Argonne National Laboratory, JL's contributions were supported by the U.S. Department of Energy, Office of Science, Office of Nuclear Physics through Contract No.~DE-AC02-06CH11357.

This research was supported in part by grant NSF PHY-2309135 to the Kavli Institute for Theoretical Physics (KITP). BSH's contributions were also supported by grant 994312 from the Simons Foundation.

TS’s contributions to this work were made possible by the Walter Benjamin Programme of the Deutsche Forschungsgemeinschaft (DFG, German Research Foundation) – 512630918. This project was also supported in part by the Black Hole Initiative at Harvard University, with support from the Gordon and Betty Moore Foundation and the John Templeton Foundation. He also acknowledges partial financial support from the Spanish Research Agency (Agencia Estatal de Investigaci\'on) through the grant IFT Centro de Excelencia Severo Ochoa No CEX2020-001007-S and PID2022-137127NB-I00 funded by MCIN/AEI/10.13039/501100011033/ FEDER, UE. This project has received funding/support from the European Union's Horizon 2020 research and innovation programme under the Marie Sklodowska-Curie Staff Exchange  grant agreement No.~101086085 -ASYMMETRY.
The opinions expressed in this publication are those of the authors and do not necessarily reflect the views of these Foundations. 

\appendix

\section{Review of the steadyon picture} \label{app:steady-review}

A natural framework for the derivation of tunneling rates is the \textit{direct approach} developed in Refs.~\cite{Andreassen:2016cff,Andreassen:2016cvx} for the decay out of a false vacuum, and developed further in Refs.~\cite{Khoury:2021zao,Steingasser:2022yqx,Chauhan:2023pur}. Following Sec.~\ref{sec:tunnel-steady}, the same arguments can be applied to represent the relevant quantities for more general states in terms of path integrals of the form
\begin{align}\label{eq:PropsScheme}
    D(x_s,\Delta t | x_* )= \int_{x(0)=x_*}^{x(\Delta t)=x_s}\mathcal{D}x\ e^{iS[x]}(\dots)\,,
\end{align}
where $D\in \{ D_F, \bar{D}_F, \underaccent{\bar}{D}_F, \bar{\underaccent{\bar}{D}}_F\}$ and the brackets represents a combination of $\delta$-functions representing their respective crossing conditions. The point $x_*$ is defined as the point within the false vacuum basin $\mathcal{F}$ beyond which the particle's wave function is localised, while $x_s$ denotes the energetically degenerate point on the other side of the barrier defining $\mathcal{F}$. See Fig.~\ref{fig:FRBarrier}.

In the special case where $x_*=x_{\rm FV}$ these integrals can be evaluated through a Wick rotation, giving rise to the familiar Euclidean-time picture.\footnote{For more details on this step, see Ref.~\cite{Andreassen:2016cvx}.} For a more general choice of $x_*$, this procedure is obstructed by the non-trivial dynamics of its asymptotic state, which are (in the semi-classical limit) dominated by an oscillatory motion. To avoid this problem while still allowing for a semi-classical evaluation of the path integral through a stationary phase approximation, the authors of Ref.~\cite{Steingasser:2024ikl} have suggested that instead of through a full Wick rotation, the theory should rather be deformed through the introduction of a small imaginary part for the Hamiltonian,
\begin{align}\label{eq:Hreg}
    H \to (1-i \epsilon) H\,,
\end{align}
with some small $\epsilon$. For time-independent systems, this is equivalent to an infinitesimal Wick rotation, but for conceptual clarity we will uphold the interpretation of a complexification of the Hamiltonian. For the action, this amounts to~\cite{Kaya:2018jdo}
\begin{align}
     S\to (1 + i \epsilon) \intd t \ \frac{m}{2}\dot{x}^2 -  (1 - i \epsilon)^2 \cdot  V (x)\,.
\end{align}
This action is extremalised by a solution $\bar{x}(t)$ of its (complex) equations of motion, 
\begin{align}
    \ddot{\bar{x}}(t)=- (1-2 i \epsilon) V'(\bar{x}(t)) \,,\label{eq:compeom}
\end{align}
with boundary conditions
\begin{gather}\label{eq:bc}
    \bar{x}(0)=x_*, \quad \bar{x}(\Delta t)=x_s\,. 
\end{gather}
Such solutions were dubbed \textit{steadyons} by the authors of Ref.~\cite{Steingasser:2024ikl}, in analogy with the \textit{instanton} solution maximising the Euclidean action describing false vacuum decay: In the complex-time plane, an instanton extends over an infinitely long Euclidean time interval, but has a vanishing extension in the real-time direction. The steadyon meanwhile is defined on the real-time axis, describing a steady loss of probability.

In order to obtain an analytical understanding of these solutions, we can make explicit their complex nature and decompose the position variable as $\bar{x}(t)\equiv x_{\rm Re}(t)+i  x_{\rm Im}(t)$. Eq.~\eqref{eq:compeom} can then be divided into its real and imaginary parts, which together can be understood as describing two real, interacting degrees of freedom,
\begin{align}
    \ddot{x}_{\rm Re} (t)= -  {\rm Re}( V'(\bar{x}(t)), \quad  \ddot{x}_{\rm Im} (t)=  \epsilon \cdot {\rm Im}( V'(\bar{x}(t))\,.
\end{align}
We first consider the case $\dot{x}_{\rm Im}(0)=0$. As the ``driving force'' acting on $x_{\rm Im}$ is suppressed by $\epsilon$, this variable itself remains negligible until large enough times $t \gtrsim \epsilon^{-1} \cdot  t_{\rm sys}$. Thus, for small enough times $t \ll \epsilon^{-1} \cdot  t_{\rm sys}$, the equation of motion for the real part $x_{\rm Re}$ reduces to the classical equation of motion. For the case of a one-dimensional point particle, this equation is solved by a periodic motion within the false vacuum basin $\mathcal{F}$. After a sufficiently long time $t \sim \epsilon^{-1} \cdot t_{\rm sys}$ the imaginary part $x_{\rm Im}$ ceases to be negligible, starting to amplify the oscillatory motion. This, in turn, drives the motion of $x_{\rm Im}$, leading to a stronger amplification of $x_{\rm Re}$. This effect gets compounded, until it ultimately allows $x_{\rm Re}$ to cross the potential barrier. Afterwards, matching the final boundary condition requires a similar dynamics to unfold, with the motion of $x_{\rm Re}$ and $x_{\rm Im}$ dampening rather than amplifying each other. As the strength of the interaction between $x_{\rm Re}$ and $x_{\rm Im}$ is controlled by $\epsilon$, the same is true for the time interval $\Delta t$ for the steadyon satisfies the boundary conditions~\eqref{eq:bc}. In Sec.~\ref{sec:tunnel-steady} $\Delta t$ emerges as an integration variable, making the result of Eq.~\eqref{eq:P-flux-master} independent of the choice of $\epsilon$ (in the appropriate limits).

\begin{figure*}[t!]
    \centering
    \includegraphics[width=0.32\textwidth]{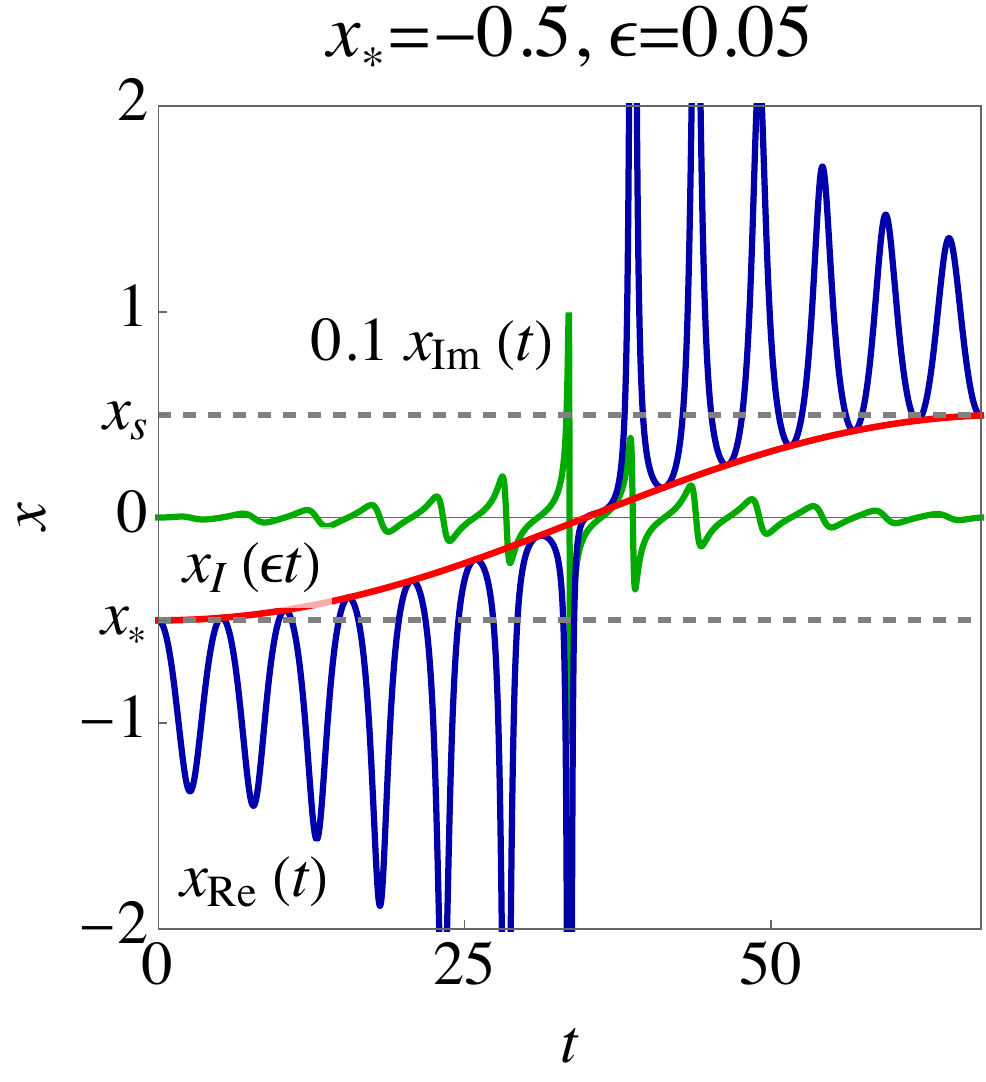}
    \includegraphics[width=0.32\textwidth]{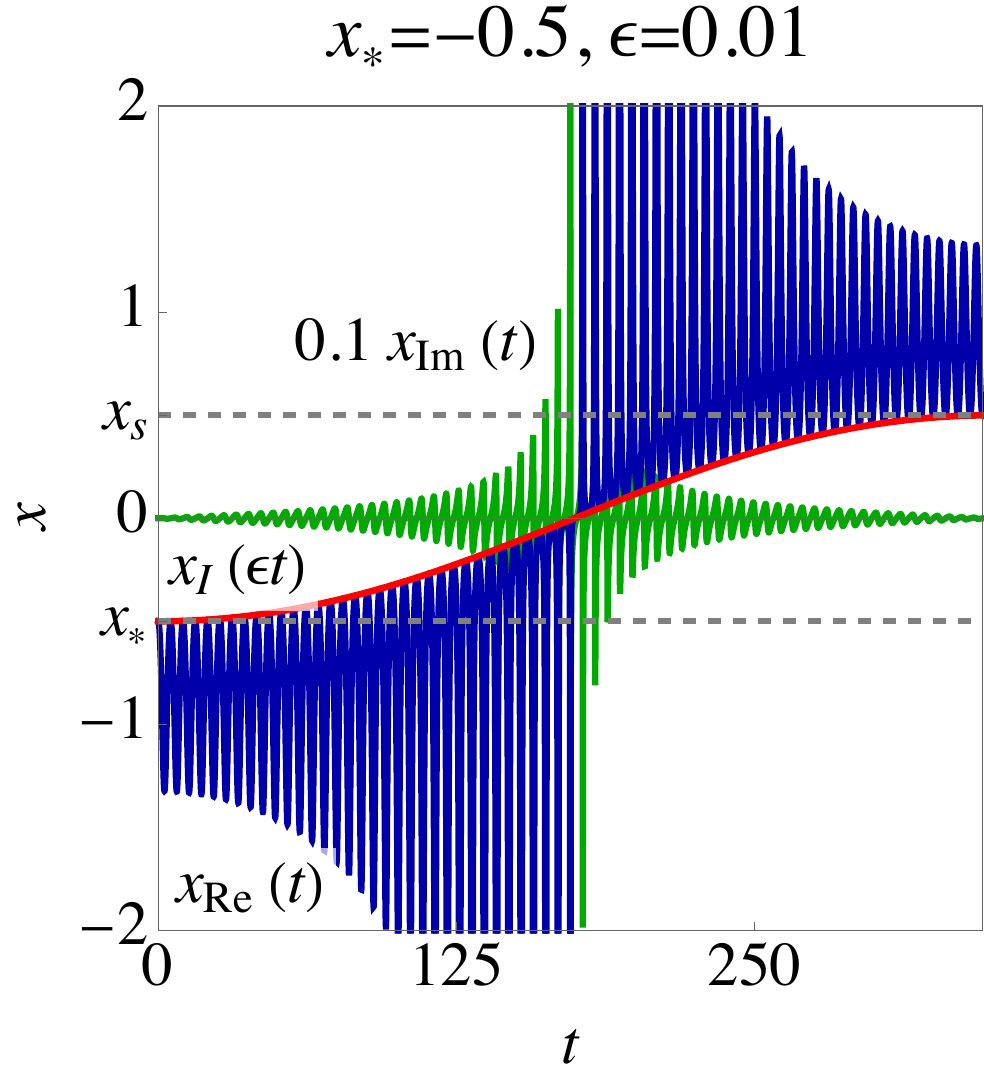}
    \includegraphics[width=0.32\textwidth]{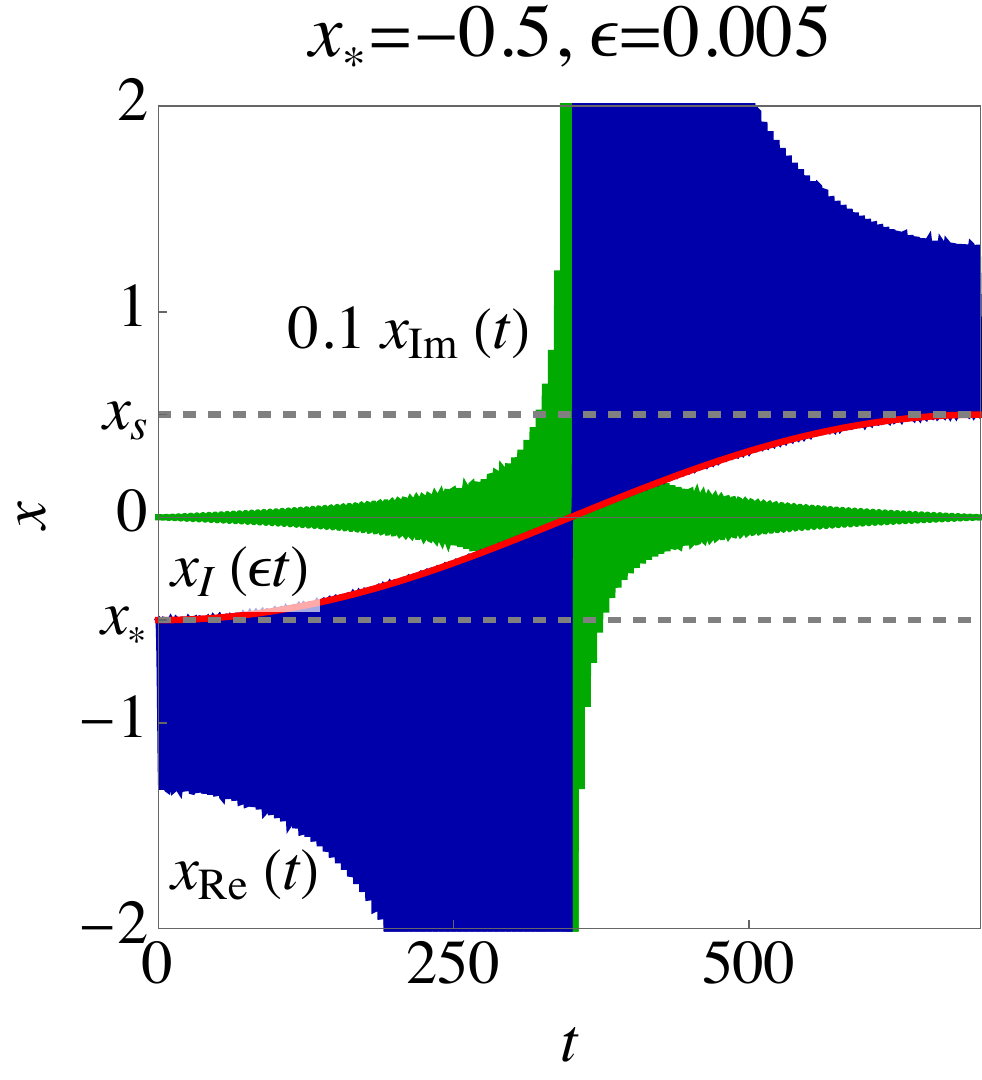}
    \caption{The periodic steadyon solution in the double well potential for $x_*=-1/2$ for different values of $\epsilon$.}
    \label{fig:ESCI}
\end{figure*}

\begin{figure*}[t!]
    \centering
    \includegraphics[width=0.32\textwidth]{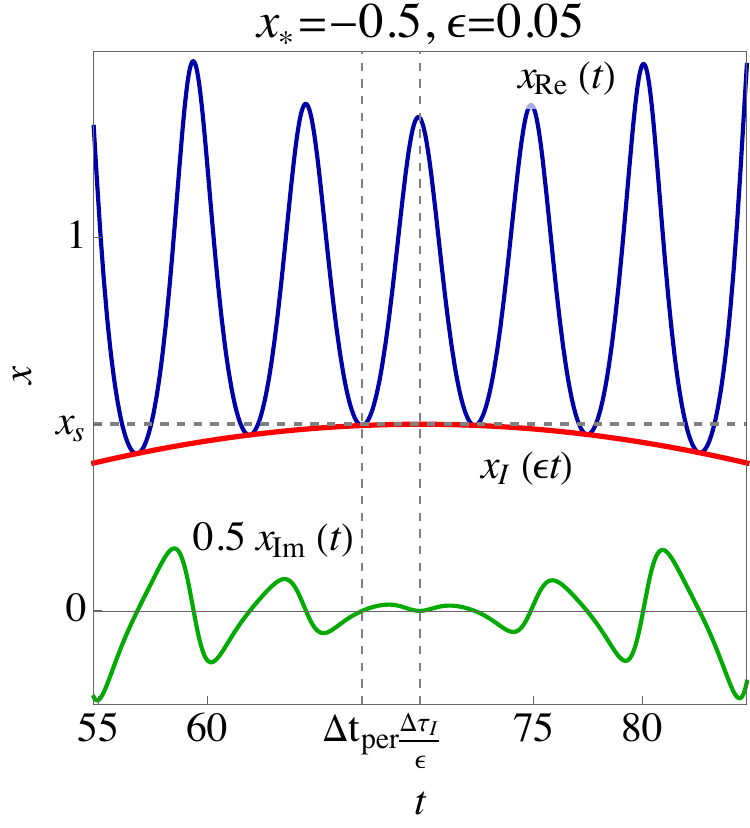}
    \includegraphics[width=0.32\textwidth]{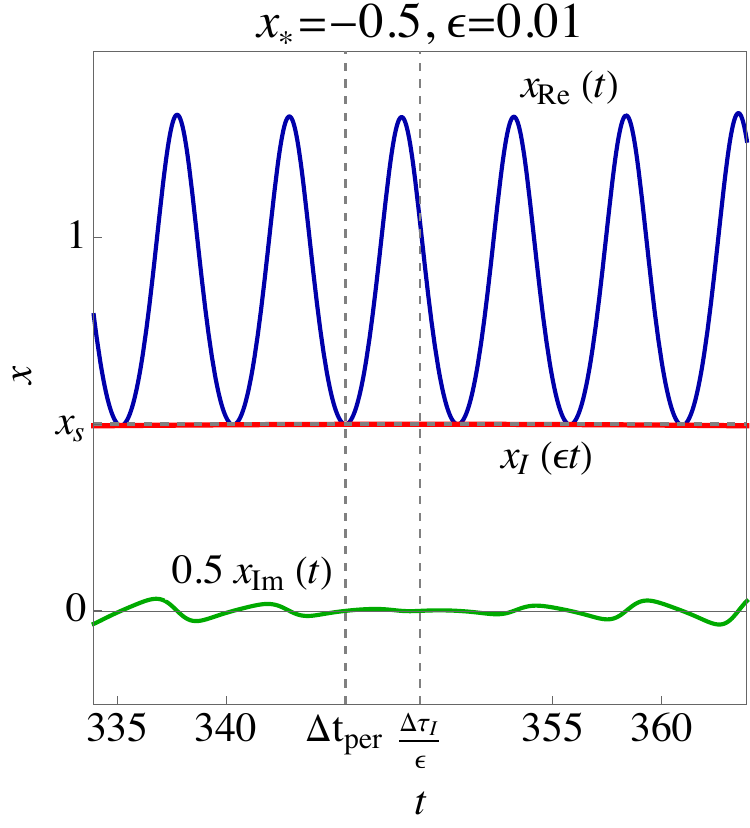}
    \includegraphics[width=0.32\textwidth]{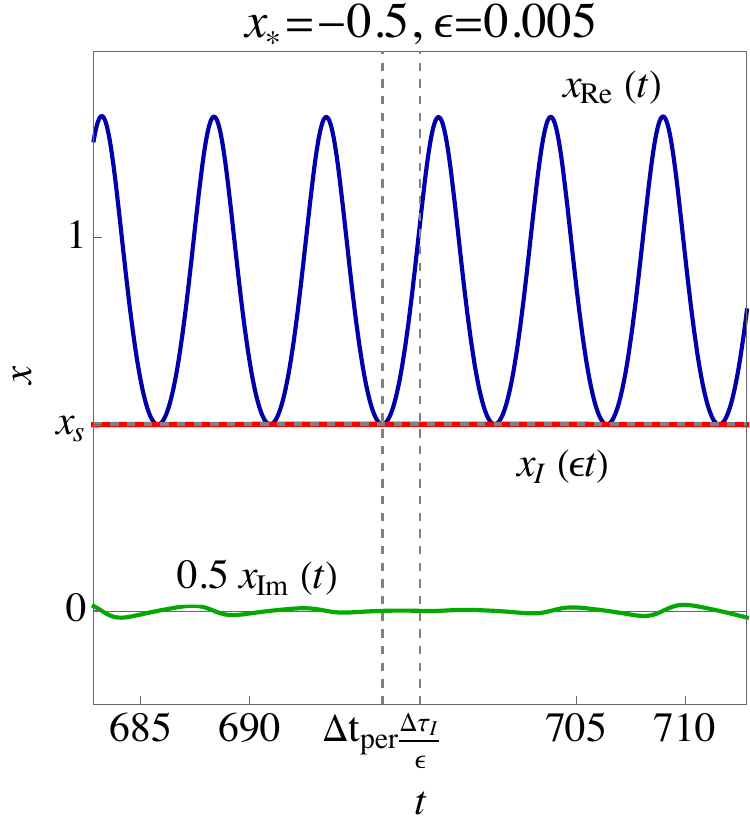}
    \caption{As Fig.~\ref{fig:ESCI}, but only for times near $t=\Delta t$ (the upper boundary of the time intervals shown in the corresponding panels of Fig.~\ref{fig:ESCI}).} 
    \label{fig:BC}
\end{figure*}
In Fig.~\ref{fig:ESCI}, we present for several choices of $\epsilon$ an example for such a dynamics in the double-well potential $V(x)= -\frac{1}{2} x^2 + \frac{1}{4} x^4$, for which Eq.~\eqref{eq:compeom} can be solved analytically. Fig.~\ref{fig:BC} meanwhile shows the same solution near the time for which the second boundary condition in~\eqref{eq:bc} is satisfied in the appropriate imits, illustrating an important subtlety of these solutions: As $x_{\rm Re}$ undergoes an oscillatory motion within the basin $\mathcal{R}$, it is not guaranteed that this variable satisfies its respective boundary condition exactly at a time when $x_{\rm Im}$ vanishes. For our analysis in Sec.~\ref{sec:tunnel-steady}, we need to be able to evaluate the propagators of interest for the interval $[\Delta t_{\rm per}-\delta t,\Delta t_{\rm per}]$. As $\Delta t= \Delta t_{\rm per}$ arises from an integral over $\Delta t$, it can always be chosen to match the boundary condition for $x_{\rm Re}$. Following our previous argument, this then amounts to an violation of the boundary condition for $x_{\rm Im}$. The discrepancy is, however, controlled by $\epsilon$. It can therefore be accounted for by a small deformation of the saddle point, ultimately leading to a subdominant change in the action suppressed by powers of $\epsilon \ll 1$. 

\begin{figure*}[t!]
    \centering
    \includegraphics[width=0.32\textwidth]{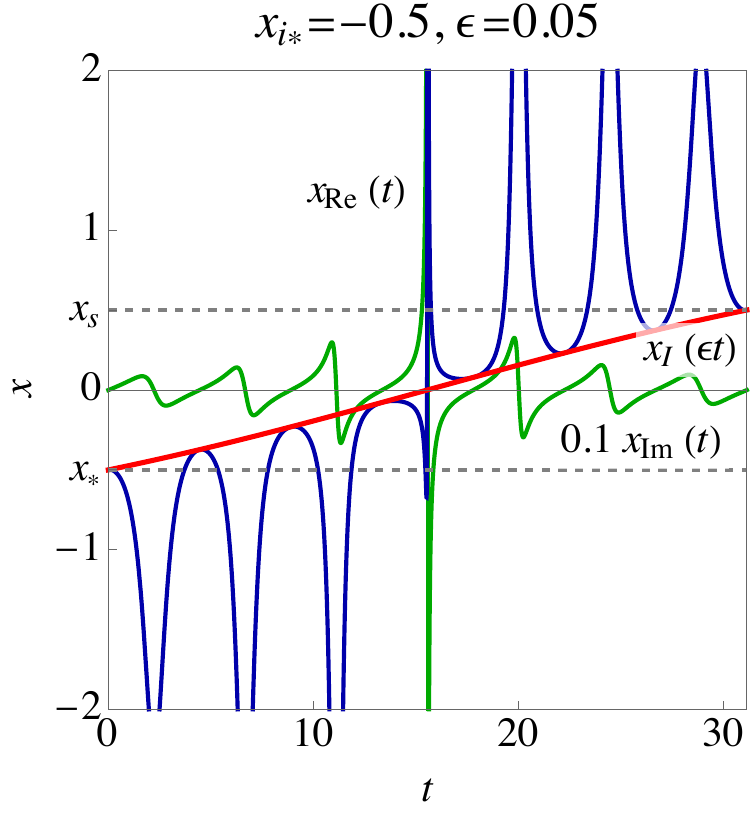}
    \includegraphics[width=0.32\textwidth]{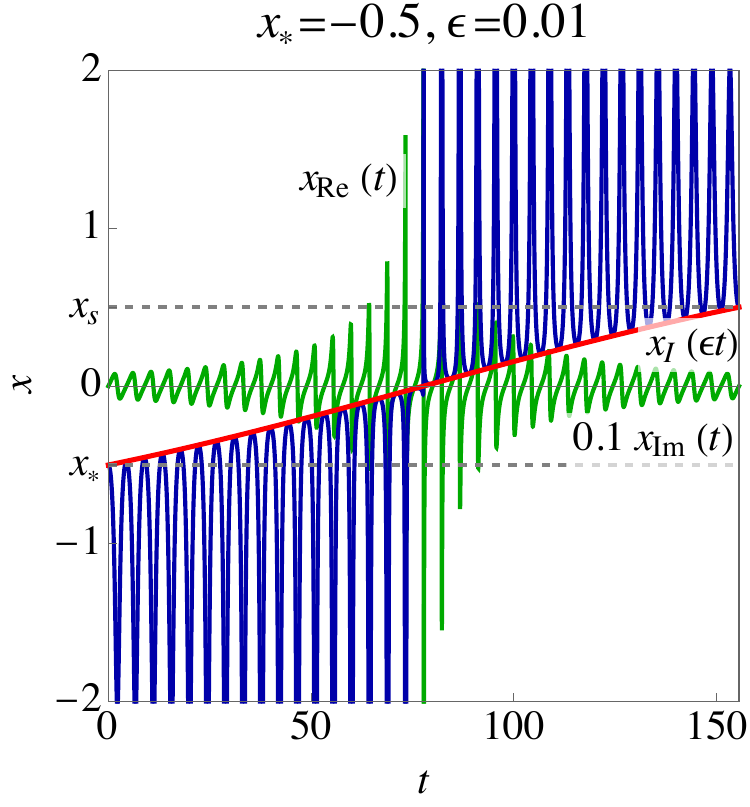}
    \includegraphics[width=0.32\textwidth]{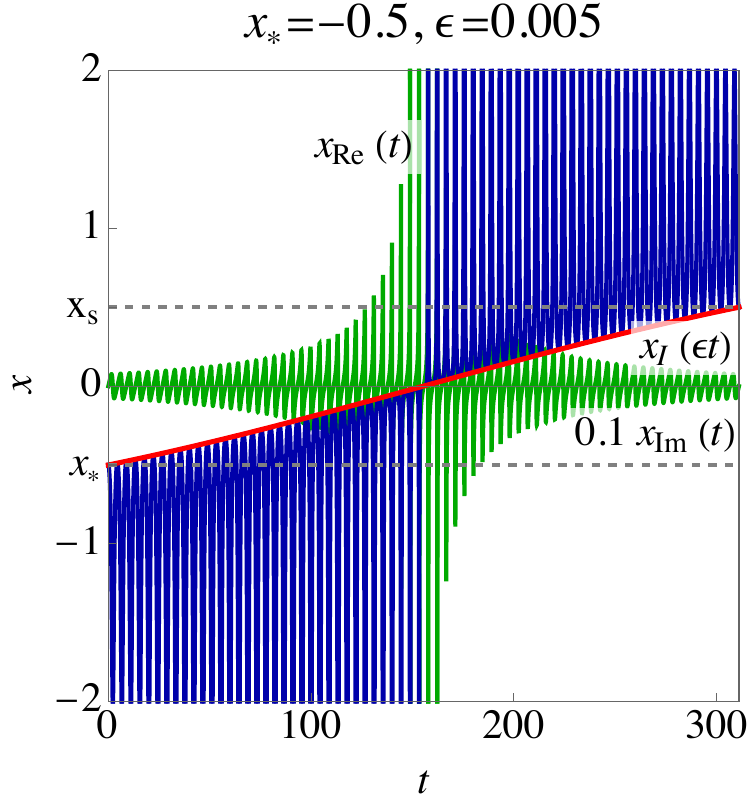}
    \caption{The non-periodic steadyon solution for $x_*=-1/2$, with $\dot{x}_{\rm Im}(0)$ chosen to match the false vacuum instanton.}
    \label{fig:ESCII}
\end{figure*}
So far, we have only considered steadyons subject to the initial boundary condition $\dot{x}_{\rm Im}(0)=0$, which we argued to represent the dominant contribution to the path integral of interest in Sec.~\ref{sec:resonance}. Figs.~\ref{fig:ESCI} and~\ref{fig:BC} also make evident why this assumption is far from obvious for more general scenarios: Without this additional assumption, the periodic steadyon describing tunneling from $x_*$ to $x_s$ could as well describe tunneling from a collection of other points within $\mathcal{F}$ to their respective energetically degenerate emergence points within $\mathcal{R}$. Conversely, for any given $x_*$ and up to offsets of order $\epsilon$, there exists a one-parameter family of steadyons. These correspond to periodic steadyons with turning points between $x_*$ and the false vacuum, but can also be characterised by their respective initial velocity of the steadyon's imaginary part $x_{\rm Im}$. See Fig.~\ref{fig:ESCII}. 

In Sec.~\ref{sec:resonance}, we used that the projection of the periodic steadyon onto the real-time axis reproduces the solution of the classical equation of motion of a point-particle confined to $\mathcal{F}$. To arrive at this conclusion, we first recall that the equation of motion for the steadyon can be understood as a Wick-rotated version of its classical counterpart controlled by $\epsilon$. Assuming the result to be an analytical function, this implies that the projection of the steadyon onto the real-time axis is a solution of the complex solution with $\epsilon=0$, which agrees with the classical equation of motion \textit{if} $x_{\rm Im}(t)=0$. The latter is indeed a valid solution along the real-time axis, but requires that $\dot{x}_{\rm Im}(t)=0$. Thus, violating this condition will lead to a deviation from the classical solution also for $x_{\rm Re}$, and thus a larger real-part of the action. A similar argument applies to the projection of the steadyon onto the imaginary-time axis $x_b(\tau)$, which solves the familiar Euclidean-time equation of motion for the instanton. This solution can therefore be understood as describing the dynamics of a point particle within the inverted potential with an initial velocity determined by $\dot{x}_{\rm Im}(0)$.

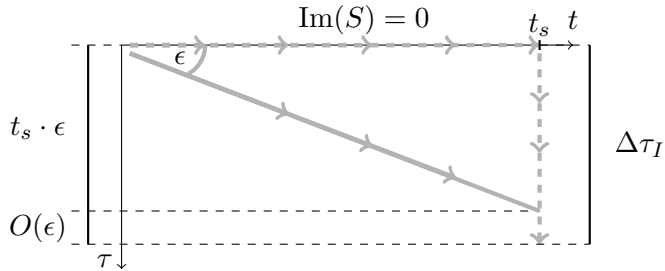
\begin{figure}[t!]
\centering
\begin{tikzpicture}[scale=2.2]
    \draw[->,gray!60, ultra thick,dashed](2.5,-0.05) -- (2.5,-0.35);
    \draw[->,gray!60, ultra thick,dashed](2.5,-0.35) -- (2.5,-0.65);
    \draw[->,gray!60, ultra thick,dashed](2.5,-0.65) -- (2.5,-1.2); 
    \draw[->](0,0)--(2.7,0); 
    \draw[<-](0,-1.35)--(0,0);
    \draw[->, gray!60, ultra thick,dashed](0.05,0)--(2.5,0);
    \draw[->, gray!60, ultra thick,dashed](0.05,0)--(0.5,0);
    \draw[->, gray!60, ultra thick,dashed](0.05,0)--(1,0);
    \draw[->, gray!60, ultra thick,dashed](0.05,0)--(1.5,0);
    \draw[->, gray!60, ultra thick,dashed](0.05,0)--(2,0);
    \draw[black, thick](-0.2,-1.2)--(-0.2,0);
    \draw[black, dashed](-0.3,0)--(0,0);
    \draw[black, thick](2.8,-1.2)--(2.8,0);
    \draw[black, dashed](2.5,0)--(2.8,0);
    \draw[black, dashed](-0.3,-1.2)--(2.8,-1.2);
    \draw[black, dashed](-0.3,-1)--(2.5,-1);
    \draw[black, thick](2.5,-0.03)--(2.5,0.03);
    \draw[gray!60,  ultra thick](2.5,-1)--(0.05,-0.05);
    \draw[->, gray!60,  ultra thick](0.05,-0.05)--(1.5,-0.61);
    \draw[->, gray!60,  ultra thick](0.05,-0.05)--(1,-0.417);
    \draw[->, gray!60,  ultra thick](0.05,-0.05)--(2,-0.8);
    \draw (2.5,0.12) node [black]{$t_s$};
    \draw (2.7,0.15) node [black]
                       {$t$};
    \draw (-0.1,-1.3) node [black]
                       {$\tau$};
    \draw (1.45,0.15) node [black]
                       {Im$\left(S\right)=0$};
    \draw (-0.5,-0.5) node [black]
                       {$t_s \cdot \epsilon$};
    \draw (-0.5,-1.1) node [black]
                       {$O(\epsilon)$};
    \draw (3.1,-0.6) node [black]{$\Delta \tau_I$};
    \draw (0.35,-0.075) node [black]
                       {$\epsilon$};
    \draw [gray!60,ultra thick](0.5,0) arc [start angle=0, end angle=-60, radius=0.2];
\end{tikzpicture}
\caption{The exponent of the decay rate arises from the integral of the steadyon's complex Lagrange function over the diagonal time contour $\gamma_\epsilon$. This contour can be deformed into pieces along the real- and imaginary-time axis, respectively. The length of the imaginary-time piece coincides with $\Delta \tau_I/2$ up to a factor of order of the regularisation parameter $\epsilon$.}
\label{fig:DeltaTauOrigin}
\end{figure}

These properties can now be used to determine the steadyon's action, which is rendered complex by the deformation~\eqref{eq:Hreg}. Relying on the interpretation of the deformation as a small Wick rotation, the action is given as a line integral over the rotated real-time axis. See Fig.~\ref{fig:dotPR}. Again assuming the steadyon to be an analytical function at least on the relevant subset of the complex-time plane, the relevant contour can be deformed into two lines on and parallel to the real- and imaginary-time axis, respectively. Following our previous discussion and focusing again on the periodic steadyon, the integral over the real-time axis coincides with the (strictly real) action of the classical steadyon in real time. Following Eq.~\eqref{eq:Pdot-res}, this part of the dynamics does not contribute to the tunneling rate, which is only sensitive to the imaginary part of the steadyon action. The latter can now be identified with the Euclidean action of the (periodic) instanton due to an additional factor of $i$ from the line element in imaginary-time direction. See Fig.~\ref{fig:DeltaTauOrigin}. Altogether, this implies that the imaginary part of the periodic steadyon's action is (in the appropriate limits) given by the Euclidean action of its corresponding periodic instanton, such that the total exponent of the decay rate reduces to the Euclidean action of the corresponding bounce,
\begin{align}
    2 \text{Im} \left( S[\bar{x}] \right) \to S_E[x_b]\,.
\end{align}

\section{Recovering the full exponential in the semiclassical approximation}\label{app:nonlinear}

In this appendix, we discuss how to directly extract the exponential $e^{-\Gamma t}$ for resonance states from $P_{\mathcal{R}}(t)$ in its long-time asymptotics. Returning to the starting point of our discussion in Sec.~\ref{sec:tunnel-steady}, and repeating our steps without taking the time derivative, and with a slightly different management of the crossing conditions, we find that the tunneling probability can be written as
\begin{align}
     P_{\mathcal{R}}(t) = \int_{\mathcal{R}} dx_f & \int dx_1 \int_0^t dt_* \, D_F(x_f,t|x_*,t_*)
     \bar{D}_F(x_*,t_*|x_1) \psi(x_1) \nonumber \\ \times & \int dx_2 \int_0^t dt_*' \, D_F^*(x_f,t|x_*,t_*') \bar{D}_F^*(x_*,t_*'|x_2) \psi^*(x_2)  \, ,
\end{align}
with no loss of generality. However, as we have already anticipated, in the semiclassical approximation one may carry out the integrals by finding the stationary phase configurations. Since the integrals over $t_*$ and $t_*'$ are complex conjugates of each other and are otherwise decoupled, the value(s) of $t_*$ and $t_*'$ for which the phase of the integrand will be stationary is the same.

Effectively, this means that the propagators appearing in this expression will (for all practical purposes in the semiclassical approximation) have the same endpoints, both in time and space. Thus, in order to establish the exponential form of the tunneling decay rate\footnote{Note that in this Appendix we have not specified for what wavefunctions it is appropriate to calculate a tunneling rate as we do in this section, i.e., by treating the tunneling process as starting solely from $x_*$. In Section~\ref{sec:resonance} we see that this is the natural identification to make for resonance states, where $x_*$ is the turning point determined by the condition $E = V(x_*)$ on the $\mathcal{F}$ side of the barrier, with $E$ the (real part of the) energy of the resonance state.}, we may simply assume that the integrals over $t_*,t_*'$ have been carried out, call their corresponding saddle points $t=0$, and study
\begin{align}
    P_{\mathcal{R}}(t) &= \int_{\mathcal{R}} dx_f \left|   D_F(x_f,t | x_*,0) \right|^2 \nonumber \\ 
    &= \int_{\mathcal{R}} dx_f \left| \int dt_0 \  D_F(x_f,t | x_s,t_0) \bar{D}_F(x_s,t_0 | x_*,0) \right|^2 \, ,
\end{align}
where we have used Eq.~\eqref{eq:Ddecomp} to decompose the propagator from $x_*$ to $x_f$.

One of the approximations we made in Sec.~\ref{sec:tunnel-steady} is that the integral over $x_f \in \mathcal{R}$ could be promoted to an integral over all of space, and that this amounted to neglecting back-tunneling contributions. While this approximation is certainly appropriate for purposes of defining and calculating the rate, it is also what makes the appearance of the exponential less obvious. If, instead of using this approximation, we had written an exact equality, we would have obtained 
\begin{align}
    P_{\mathcal{R}}(t) = &\int dx_f \left| \int dt_0 \  D_F(x_f,t|x_s,t_0) \bar{D}_F(x_s,t_0|x_*,0)\right|^2 \nonumber \\
    - &\int_{\mathcal{F}} dx_f \left| \int dt_0 \  D_F(x_f,t|x_s,t_0) \bar{D}_F(x_s,t_0|x_*,0) \right|^2  \, ,
\end{align}
which contains a second term, containing the contributions from paths including back-tunneling paths that we previously neglected.

We thus anticpate it to contain the remaining terms necessary to restore the full exponential dependence. Inserting a second crossing condition, this time \textit{back} to the false vacuum, we get 
\begin{align}
    P_{\mathcal{R}}(t) = &\int dx_f \left| \int dt_0 D_F(x_f,t|x_s,t_0) \bar{D}_F(x_s,t_0|x_*,0)  \right|^2 \nonumber \\
    - &\int_{\mathcal{F}} dx_f \left| \int_{t_0'>t_0} dt_0 dt_0' D_F(x_f,t|x_*,t_0') \bar{D}_F(x_*,t_0'|x_s,t_0) \bar{D}_F(x_s,t_0|x_*,0)  \right|^2  \nonumber \\
    = &\int dx_f \bigg( \left| \int dt_0 D_F(x_f,t|x_s,t_0) \bar{D}_F(x_s,t_0|x_*,0)  \right|^2 \nonumber \\
    & \quad\quad - \left| \int_{t_0'>t_0} dt_0 dt_0' D_F(x_f,t|x_*,t_0') \bar{D}_F(x_*,t_0'|x_s,t_0) \bar{D}_F(x_s,t_0|x_*,0) \right|^2 \bigg) \nonumber \\
    + &\int_{\mathcal{R}} dx_f \left| \int_{t_0'>t_0} dt_0 dt_0' D_F(x_f,t|x_*,t_0')  \bar{D}_F(x_*,t_0'|x_s,t_0) \bar{D}_F(x_s,t_0|x_*,0) \right|^2 \, .
\end{align}
The structure that emerges is clear: one can reconstruct the probability in terms of integrals over the real line by adding successive crossing conditions. The resulting series is 
\begin{align}
    P_{\mathcal{R}}(t) &= \int dx_f \sum_{n=0}^\infty \Bigg| \int_{t > t_{2n+1} > \ldots > t_1 > t_0 = 0 } \!\!\!\!\!\!\!\!\!\!\!\!\!\!\!\!\!\!\!\!\!\!\!\!\!\!\!\!\!\!\!\!\!\!\!\!\!\!\!\!\! dt_1 \ldots dt_{2n+1} \quad   D_F(x_f,t|x_s,t_{2n+1}) \bar{D}_F(x_s,t_{2n+1}|x_*,t_{2n})  \nonumber \\ & \quad\quad\quad\quad\quad\quad\quad\quad\quad\quad\quad\quad\quad \times \left( \prod_{i=0}^{n-1} \bar{D}_F(x_*,t_{2i+2}|x_s,t_{2i+1})  \bar{D}_F(x_s,t_{2i+1}|x_*,t_{2i}) \right)  \Bigg|^2 \nonumber \\
    & - \int dx_f \sum_{n=1}^\infty \Bigg| \int_{t > t_{2n} > \ldots > t_1 > t_0 = 0 } \!\!\!\!\!\!\!\!\!\!\!\!\!\!\!\!\!\!\!\!\!\!\!\!\!\!\!\!\!\!\!\!\!\!\!\! dt_1 \ldots dt_{2n} \quad D_F(x_f,t|x_*,t_{2n}) \nonumber \\ & \quad\quad\quad\quad\quad\quad\quad\quad\quad\quad\quad\quad\quad \times \left( \prod_{i=0}^{n-1}  \bar{D}_F(x_*,t_{2i+2}|x_s,t_{2i+1}) \bar{D}_F(x_s,t_{2i+1}|x_*,t_{2i}) \right)    \Bigg|^2 \, ,
\end{align}
which describes the probability to have tunneled out of the false vacuum via an arbitrary number of intermediate tunneling processes into and out of itself.

Furthermore, it turns out that the integral over $x_f$ may actually be done analytically -- in fact, as we did in Sec.~\ref{sec:tunnel-steady}. Note that $x_f$ will only appear through the two propagators at the very end of the tunneling sequence, leading in one case to the integral 
\begin{align}
    &\int dx_f D(x_f,t|x_s,t_{2n+1}) D^*(x_f,t|x_s,t_{2n+1}') \\
    &= \theta(t_{2n+1}' - t_{2n+1}) D(x_s,t_{2n+1}'|x_s,t_{2n+1}) + \theta(t_{2n+1}-t_{2n+1}') D^*(x_s,t_{2n+1}|x_s,t_{2n+1}')  \, , \nonumber
\end{align}
and in the other case an equivalent integral where the triplet $(x_s,t_{2n+1},t_{2n+1}')$ is replaced by $(x_*,t_{2n},t_{2n}')$. To be explicit, this leads to 
\begin{align}
   & P_{\mathcal{R}}(t) = \nonumber \\ 
   & \quad \sum_{n=0}^\infty \int_{\substack{ t > t_{2n+1}' > \ldots > t_1' > t_0 = 0 \\ t_{2n+1}' > t_{2n+1} > \ldots > t_1 > t_0 = 0 }}  dt_1 \ldots dt_{2n+1} \, dt_1' \ldots dt_{2n+1}' \nonumber \\
    & \quad\quad\quad \times D(x_s,t_{2n+1}'|x_s,t_{2n+1}) \bar{D}(x_s,t_{2n+1}|x_*,t_{2n}) \bar{D}^*(x_s,t_{2n+1}'|x_*,t_{2n}')  \nonumber \\
    & \quad\quad\quad \times \left( \prod_{i=0}^{n-1}  \bar{D}(x_*,t_{2i+2}|x_s,t_{2i+1}) \bar{D}(x_s,t_{2i+1}|x_*,t_{2i}) \bar{D}^*(x_*,t_{2i+2}'|x_s,t_{2i+1}') \bar{D}^*(x_s,t_{2i+1}'|x_*,t_{2i}') \right)  \nonumber \\
    & - \sum_{n=1}^\infty \int_{\substack{ t > t_{2n}' > \ldots > t_1' > t_0 = 0 \\ t_{2n}' > t_{2n} > \ldots > t_1 > t_0 = 0 }}  dt_1 \ldots dt_{2n} \, dt_1' \ldots dt_{2n}' \nonumber \\
    & \quad\quad\quad \times D(x_*,t_{2n}'|x_*,t_{2n}) \nonumber \\
    & \quad\quad\quad \times \left( \prod_{i=0}^{n-1}  \bar{D}(x_*,t_{2i+2}|x_s,t_{2i+1}) \bar{D}(x_s,t_{2i+1}|x_*,t_{2i}) \bar{D}^*(x_*,t_{2i+2}'|x_s,t_{2i+1}') \bar{D}^*(x_s,t_{2i+1}'|x_*,t_{2i}')  \right)  \nonumber \\
    & + {\rm c.c.} \, .
\end{align}
Finally, one can simplify these expressions further by carrying out the integrals over $t_{2n+1}$ and $t_{2n}$, respectively. The result is
\begin{align}
    & P_{\mathcal{R}}(t) = \nonumber \\ & \quad \sum_{n=0}^\infty \int_{\substack{ t > t_{2n+1}' > \ldots > t_1' > t_0 = 0 \\ t_{2n+1}' > t_{2n} > \ldots > t_1 > t_0 = 0 }}  dt_1 \ldots dt_{2n} \, dt_1' \ldots dt_{2n+1}' \nonumber \\
    & \quad\quad\quad \times \bar{D}^*(x_s,t_{2n+1}'|x_*,t_{2n}') D(x_s,t_{2n+1}'|x_*,t_{2n})  \nonumber \\
    & \quad\quad\quad \times \left( \prod_{i=0}^{n-1} \bar{D}(x_*,t_{2i+2}|x_s,t_{2i+1}) \bar{D}(b,t_{2i+1}|x_*,t_{2i})  \bar{D}^*(x_*,t_{2i+2}'|x_s,t_{2i+1}')  \bar{D}^*(x_s,t_{2i+1}'|x_*,t_{2i}') \right)  \nonumber \\
    & - \sum_{n=1}^\infty \int_{\substack{ t > t_{2n}' > \ldots > t_1' > t_0 = 0 \\ t_{2n}' > t_{2n-1} > \ldots > t_1 > t_0 = 0 }}  dt_1 \ldots dt_{2n-1} \, dt_1' \ldots dt_{2n}' \nonumber \\
    & \quad\quad\quad \times D(x_*,t_{2n}'|x_s,t_{2n-1}) \bar{D}(x_s,t_{2n-1}|x_*,t_{2n-2})  \nonumber \\
    & \quad\quad\quad \times \left( \prod_{i=0}^{n-2}  \bar{D}(x_*,t_{2i+2}|x_s,t_{2i+1}) \bar{D}(b,t_{2i+1}|x_*,t_{2i})  \bar{D}^*(x_*,t_{2i+2}'|x_s,t_{2i+1}') \bar{D}^*(x_*,t_{2i+1}'|x_*,t_{2i}') \right)  \nonumber \\
    & + {\rm c.c.} \, .
\end{align}
So far, these are very big expressions with no further simplification in sight with only formal manipulations.
Therefore, it is highly beneficial at this point to evaluate the propagators $D$, $\bar{D}$ in the semiclassical approximation. Concretely, using that 
\begin{align}
    \bar{D}(x_*,t_0'|x_s,t_0) \approx D(x_*,t_0'|x_s,t_0) \approx A \exp \left( i S(x_*,x_s,t_0'-t_0) \right) \, ,
\end{align}
where $S(x_*,x_s,\Delta t)$ is the (complex) action corresponding to the trajectory of the particle between points $x_s$ and $x_*$ in a time interval $\Delta t$ that extremises said action, and $A$ is some $\mathcal{O}(1)$ prefactor (not exponentially enhanced or suppressed by $\hbar$). Note that $S(x_*,x_s,\Delta t) = S(x_s,x_*,\Delta t)$ because the extremal trajectories may be reparametrised to run backwards and yield the same action.
We may profit from this approximation because in the limit of large $\Delta t$, the action takes the form (see Appendix~\ref{app:steady-review}) 
\begin{equation}
    S(x_*,x_s,\Delta t) \approx \Delta t \, S_r(x_*,x_s) + i S_E(x_*,x_s)/2 \, . \label{eq:S-steady-approx}
\end{equation}
With these identifications, in the long time limit of the integrals over $t_i, t_i', \ldots$, the products of propagators in the previous equation simplify dramatically to
\begin{align}
    & P_{\mathcal{R}}(t) = \nonumber \\ & \quad \sum_{n=0}^\infty \int_{\substack{ t > t_{2n+1}' > \ldots > t_1' > t_0 = 0 \\ t_{2n+1}' > t_{2n} > \ldots > t_1 > t_0 = 0 }}  dt_1 \ldots dt_{2n} \, dt_1' \ldots dt_{2n+1}' \, A^{4n+1} \nonumber \\ 
    & \quad\quad \times \exp \left( i t_{2n+1}' S_r(x_*,x_s) - (2n+1) S_E(x_*,x_s)/2 \right)  \exp \left( - i t_{2n+1}' S_r(x_*,x_s) - (2n+1) S_E(x_*,x_s)/2 \right) \nonumber \\
    & - \sum_{n=1}^\infty \int_{\substack{ t > t_{2n}' > \ldots > t_1' > t_0 = 0 \\ t_{2n}' > t_{2n-1} > \ldots > t_1 > t_0 = 0 }}  dt_1 \ldots dt_{2n-1} \, dt_1' \ldots dt_{2n}'  \, A^{4n-1} \nonumber \\ 
    & \quad\quad\quad \times \exp \left( i t_{2n}' S_r(x_*,x_s) - 2n S_E(x_*,x_s)/2 \right)  \exp \left( - i t_{2n}' S_r(x_*,x_s) - 2n S_E(x_*,x_s)/2 \right) \nonumber \\
    & + {\rm c.c.} \, ,
\end{align}
and then finally
\begin{align}
     P_{\mathcal{R}}(t) &= \sum_{n=1}^\infty \frac{(-1)^{n+1}}{n!} (A t)^n e^{- n S_E(x_*,x_s)} \nonumber \\
    &= 1 - \exp \left( - t A e^{- S_E(x_*,x_s)} \right) \, .
\end{align}
It follows that the probability of being in the region $\mathcal{R}$ behaves as expected, fully determined by the anticipated rate
\begin{equation}\label{eq:antic}
    \Gamma = A \, e^{- S_E(x_*,x_s)} \, .
\end{equation}
Note that this formula is inherently asymmetric: tunneling occurs out of region $\mathcal{F}$ into region $\mathcal{R}$. In the context of this derivation, this happens by design: the first propagator starts from $x_*$ and it is \textit{enforced} that the path ends in $\mathcal{R}$, thus the only possibility is that a tunneling event occurs. The full dynamics (in the semiclassical approximation) is recovered after convoluting this with the dynamics of the wavefunction in $\mathcal{F}$. If the extent of the region $\mathcal{R}$ is infinite, it follows that eventually all probability will be depleted from $\mathcal{F}$. If the extent of the region $\mathcal{R}$ is not infinite, there will eventually be back-tunneling processes that are sensitive to the details of the potential past the barrier, which are not captured by the assumption~\eqref{eq:S-steady-approx} following from steadyon contributions to the propagator. Finally, we note that, if, contrary to our working assumption, one happened to choose $\mathcal{F}$ to be the lowest energy sector of the potential, the above derivation does not follow through for points $x_*$ that do not have a corresponding point $x_s$ with equal potential energy at the other side of the barrier to define a crossing condition, because there will be no semiclassical stationary phase configurations that contribute.

\bibliographystyle{JHEP}
\bibliography{main.bib}

@article{Andreassen:2016cff,
    author = "Andreassen, Anders and Farhi, David and Frost, William and Schwartz, Matthew D.",
    title = "{Direct Approach to Quantum Tunneling}",
    eprint = "1602.01102",
    archivePrefix = "arXiv",
    primaryClass = "hep-th",
    doi = "10.1103/PhysRevLett.117.231601",
    journal = "Phys. Rev. Lett.",
    volume = "117",
    number = "23",
    pages = "231601",
    year = "2016"
}

@article{Andreassen:2016cvx,
    author = "Andreassen, Anders and Farhi, David and Frost, William and Schwartz, Matthew D.",
    title = "{Precision decay rate calculations in quantum field theory}",
    eprint = "1604.06090",
    archivePrefix = "arXiv",
    primaryClass = "hep-th",
    doi = "10.1103/PhysRevD.95.085011",
    journal = "Phys. Rev. D",
    volume = "95",
    number = "8",
    pages = "085011",
    year = "2017"
}

@article{DORMAND198019,
title = {A family of embedded Runge-Kutta formulae},
journal = {Journal of Computational and Applied Mathematics},
volume = {6},
number = {1},
pages = {19-26},
year = {1980},
issn = {0377-0427},
doi = {https://doi.org/10.1016/0771-050X(80)90013-3},
url = {https://www.sciencedirect.com/science/article/pii/0771050X80900133},
author = {J.R. Dormand and P.J. Prince}
}

@article{Khoury:2021zao,
    author = "Khoury, Justin and Steingasser, Thomas",
    title = "{Gauge hierarchy from electroweak vacuum metastability}",
    eprint = "2108.09315",
    archivePrefix = "arXiv",
    primaryClass = "hep-ph",
    doi = "10.1103/PhysRevD.105.055031",
    journal = "Phys. Rev. D",
    volume = "105",
    number = "5",
    pages = "055031",
    year = "2022"
}

@phdthesis{Steingasser:2022yqx,
    author = "Steingasser, Thomas",
    title = "{New perspectives on solitons and instantons in the Standard Model and beyond}",
    doi = "10.5282/edoc.30495",
    school = "Munich U.",
    year = "2022"
}

@article{Chauhan:2023pur,
    author = "Chauhan, Garv and Steingasser, Thomas",
    title = "{Gravity-improved metastability bounds for the Type-I seesaw mechanism}",
    eprint = "2304.08542",
    archivePrefix = "arXiv",
    primaryClass = "hep-ph",
    reportNumber = "MIT-CTP/5520",
    doi = "10.1007/JHEP09(2023)151",
    journal = "JHEP",
    volume = "09",
    pages = "151",
    year = "2023"
}

@article{Steingasser:2024ikl,
    author = "Steingasser, Thomas and Kaiser, David I.",
    title = "{Toward quantum tunneling from excited states: Recovering imaginary-time instantons from a real-time analysis}",
    eprint = "2402.00099",
    archivePrefix = "arXiv",
    primaryClass = "hep-th",
    reportNumber = "MIT-CTP/5672",
    doi = "10.1103/PhysRevD.111.096009",
    journal = "Phys. Rev. D",
    volume = "111",
    number = "9",
    pages = "096009",
    year = "2025"
}

@article{Steingasser:2023gde,
    author = {Steingasser, Thomas and K{\"o}nig, Morgane and Kaiser, David I.},
    title = "{Finite-temperature instantons from first principles}",
    eprint = "2310.19865",
    archivePrefix = "arXiv",
    primaryClass = "hep-th",
    reportNumber = "MIT-CTP/5638",
    doi = "10.1103/PhysRevD.110.L111902",
    journal = "Phys. Rev. D",
    volume = "110",
    number = "11",
    pages = "L111902",
    year = "2024"
}

@inproceedings{Lin:2007iq,
    author = "Lin, Huey-Wen and Cohen, Saul D.",
    title = "{Lattice QCD beyond ground states}",
    booktitle = "{4th International Workshop on Numerical Analysis and Lattice QCD}",
    eprint = "0709.1902",
    archivePrefix = "arXiv",
    primaryClass = "hep-lat",
    reportNumber = "JLAB-THY-07-720",
    month = "9",
    year = "2007"
}

@article{Wagman:2024rid,
    author = "Wagman, Michael L.",
    title = "{Lanczos Algorithm, the Transfer Matrix, and the Signal-to-Noise Problem}",
    eprint = "2406.20009",
    archivePrefix = "arXiv",
    primaryClass = "hep-lat",
    reportNumber = "FERMILAB-PUB-24-0320-T",
    doi = "10.1103/pcvc-734h",
    journal = "Phys. Rev. Lett.",
    volume = "134",
    number = "24",
    pages = "241901",
    year = "2025"
}

@article{Fox:1981xz,
    author = "Fox, G. and Gupta, R. and Martin, O. and Otto, S.",
    title = "{Monte Carlo Estimates of the Mass Gap of the O(2) and O(3) Spin Models in (1+1)-dimensions}",
    reportNumber = "CALT-68-866",
    doi = "10.1016/0550-3213(82)90384-4",
    journal = "Nucl. Phys. B",
    volume = "205",
    pages = "188--220",
    year = "1982"
}

@book{Gattringer:2010zz,
    author = "Gattringer, Christof and Lang, Christian B.",
    title = "{Quantum chromodynamics on the lattice}",
    doi = "10.1007/978-3-642-01850-3",
    isbn = "978-3-642-01849-7, 978-3-642-01850-3",
    publisher = "Springer",
    address = "Berlin",
    volume = "788",
    year = "2010"
}

@book{Montvay:1994cy,
    author = "Montvay, I. and Munster, G.",
    title = "{Quantum fields on a lattice}",
    doi = "10.1017/CBO9780511470783",
    isbn = "978-0-521-59917-7, 978-0-511-87919-7",
    publisher = "Cambridge University Press",
    series = "Cambridge Monographs on Mathematical Physics",
    month = "3",
    year = "1997"
}

@article{Blossier:2009kd,
    author = "Blossier, Benoit and Della Morte, Michele and von Hippel, Georg and Mendes, Tereza and Sommer, Rainer",
    title = "{On the generalized eigenvalue method for energies and matrix elements in lattice field theory}",
    eprint = "0902.1265",
    archivePrefix = "arXiv",
    primaryClass = "hep-lat",
    reportNumber = "DESY-09-014, SFB-CPP-09-10, MKPH-T-09-01, LPT-ORSAY-09-05",
    doi = "10.1088/1126-6708/2009/04/094",
    journal = "JHEP",
    volume = "04",
    pages = "094",
    year = "2009"
}

@article{Wentzel:1926WKB,  
  author       = {Wentzel, Gregor},  
  title        = {Eine Verallgemeinerung der Quantenbedingungen für die Zwecke der Wellenmechanik},  
  journal      = {Zeitschrift für Physik},  
  volume       = {38},  
  number       = {6--7},  
  pages        = {518--529},  
  year         = {1926}
}

@article{Hund:1927Tunnel,  
  author       = {Hund, Friedrich},  
  title        = {Zur Deutung der Molekelspektren. III},  
  journal      = {Zeitschrift für Physik},  
  volume       = {43},  
  pages        = {805--826},  
  year         = {1927}
}

@article{Gamow:1928AlphaDecay,  
  author       = {Gamow, G.},  
  title        = {Zur Quantentheorie des Atomkernes},  
  journal      = {Zeitschrift für Physik},  
  volume       = {51},  
  pages        = {204--212},  
  year         = {1928}
}

@article{Coleman:1977py,
    author = "Coleman, Sidney R.",
    title = "{The Fate of the False Vacuum. 1. Semiclassical Theory}",
    reportNumber = "HUTP-77-A004",
    doi = "10.1103/PhysRevD.16.1248",
    journal = "Phys. Rev. D",
    volume = "15",
    pages = "2929--2936",
    year = "1977",
    note = "[Erratum: Phys.Rev.D 16, 1248 (1977)]"
}

@article{Callan:1977pt,
    author = "Callan, Jr., Curtis G. and Coleman, Sidney R.",
    title = "{The Fate of the False Vacuum. 2. First Quantum Corrections}",
    reportNumber = "HUTP-77-A032",
    doi = "10.1103/PhysRevD.16.1762",
    journal = "Phys. Rev. D",
    volume = "16",
    pages = "1762--1768",
    year = "1977"
}

@article{Kobzarev:1974cp,
    author = "Kobzarev, I. Yu. and Okun, L. B. and Voloshin, M. B.",
    title = "{Bubbles in Metastable Vacuum}",
    reportNumber = "ITEP-81-1974",
    journal = "Yad. Fiz.",
    volume = "20",
    pages = "1229--1234",
    year = "1974"
}

@article{Devoto:2022qen,
    author = "Devoto, Federica and Devoto, Simone and Di Luzio, Luca and Ridolfi, Giovanni",
    title = "{False vacuum decay: an introductory review}",
    eprint = "2205.03140",
    archivePrefix = "arXiv",
    primaryClass = "hep-ph",
    doi = "10.1088/1361-6471/ac7f24",
    journal = "J. Phys. G",
    volume = "49",
    number = "10",
    pages = "103001",
    year = "2022"
}

@article{GurneyCondon:1928Nature,  
  author       = {Gurney, R. W. and Condon, E. U.},  
  title        = {Wave Mechanics and Radioactive Disintegration},  
  journal      = {Nature},  
  volume       = {122},  
  pages        = {439},  
  year         = {1928},  
  note         = {Announcement of tunneling explanation for alpha-decay},  
}

@article{GurneyCondon:1929PR,  
  author       = {Gurney, R. W. and Condon, E. U.},  
  title        = {Quantum Mechanics and Radioactive Disintegration},  
  journal      = {Physical Review},  
  volume       = {33},  
  pages        = {127--140},  
  year         = {1929},  
  note         = {Detailed quantitative treatment relating half-lives to barrier penetration},  
}

@article{MacColl:1932WavePackets,  
  author       = {MacColl, L. A.},  
  title        = {Note on the Transmission and Reflection of Wave Packets by Potential Barriers},  
  journal      = {Physical Review},  
  volume       = {40},  
  pages        = {621--626},  
  year         = {1932},  
  note         = {First clear time-domain analysis of tunneling (tunneling time origin)},  
}

@article{Zinn-Justin:2002ecy,
    author = "Zinn-Justin, Jean",
    title = "{Quantum field theory and critical phenomena}",
    journal = "Int. Ser. Monogr. Phys.",
    volume = "113",
    pages = "1--1054",
    year = "2002"
}

@article{Witten:2010cx,
    author = "Witten, Edward",
    editor = "Andersen, Joergen E. and Boden, Hans U. and Hahn, Atle and Himpel, Benjamin",
    title = "{Analytic Continuation Of Chern-Simons Theory}",
    eprint = "1001.2933",
    archivePrefix = "arXiv",
    primaryClass = "hep-th",
    journal = "AMS/IP Stud. Adv. Math.",
    volume = "50",
    pages = "347--446",
    year = "2011"
}

@article{Tanizaki:2014xba,
    author = "Tanizaki, Yuya and Koike, Takayuki",
    title = "{Real-time Feynman path integral with Picard{\textendash}Lefschetz theory and its applications to quantum tunneling}",
    eprint = "1406.2386",
    archivePrefix = "arXiv",
    primaryClass = "math-ph",
    reportNumber = "RIKEN-QHP-156",
    doi = "10.1016/j.aop.2014.09.003",
    journal = "Annals Phys.",
    volume = "351",
    pages = "250--274",
    year = "2014"
}

@article{Cherman:2014sba,
    author = "Cherman, Aleksey and Unsal, Mithat",
    title = "{Real-Time Feynman Path Integral Realization of Instantons}",
    eprint = "1408.0012",
    archivePrefix = "arXiv",
    primaryClass = "hep-th",
    reportNumber = "FTPI-MINN-14-18, UMN-TH-3343-14",
    month = "7",
    year = "2014"
}

@article{Dunne:2015eaa,
    author = {Dunne, Gerald V. and {\"U}nsal, Mithat},
    title = "{What is QFT? Resurgent trans-series, Lefschetz thimbles, and new exact saddles}",
    eprint = "1511.05977",
    archivePrefix = "arXiv",
    primaryClass = "hep-lat",
    doi = "10.22323/1.251.0010",
    journal = "PoS",
    volume = "LATTICE2015",
    pages = "010",
    year = "2016"
}

@article{Bramberger:2016yog,
    author = "Bramberger, Sebastian F. and Lavrelashvili, George and Lehners, Jean-Luc",
    title = "{Quantum tunneling from paths in complex time}",
    eprint = "1605.02751",
    archivePrefix = "arXiv",
    primaryClass = "hep-th",
    doi = "10.1103/PhysRevD.94.064032",
    journal = "Phys. Rev. D",
    volume = "94",
    number = "6",
    pages = "064032",
    year = "2016"
}

@article{Michel:2019nwa,
    author = "Michel, Florent",
    title = "{Parametrized Path Approach to Vacuum Decay}",
    eprint = "1911.12765",
    archivePrefix = "arXiv",
    primaryClass = "quant-ph",
    doi = "10.1103/PhysRevD.101.045021",
    journal = "Phys. Rev. D",
    volume = "101",
    number = "4",
    pages = "045021",
    year = "2020"
}

@article{Mou:2019gyl,
    author = "Mou, Zong-Gang and Saffin, Paul M. and Tranberg, Anders",
    title = "{Quantum tunnelling, real-time dynamics and Picard-Lefschetz thimbles}",
    eprint = "1909.02488",
    archivePrefix = "arXiv",
    primaryClass = "hep-th",
    doi = "10.1007/JHEP11(2019)135",
    journal = "JHEP",
    volume = "11",
    pages = "135",
    year = "2019"
}

@article{Hertzberg:2019wgx,
    author = "Hertzberg, Mark P. and Yamada, Masaki",
    title = "{Vacuum Decay in Real Time and Imaginary Time Formalisms}",
    eprint = "1904.08565",
    archivePrefix = "arXiv",
    primaryClass = "hep-th",
    doi = "10.1103/PhysRevD.100.016011",
    journal = "Phys. Rev. D",
    volume = "100",
    number = "1",
    pages = "016011",
    year = "2019"
}

@article{Ai:2019fri,
    author = {Ai, Wen-Yuan and Garbrecht, Bj{\"o}rn and Tamarit, Carlos},
    title = "{Functional methods for false vacuum decay in real time}",
    eprint = "1905.04236",
    archivePrefix = "arXiv",
    primaryClass = "hep-th",
    reportNumber = "TUM-HEP-1201-19",
    doi = "10.1007/JHEP12(2019)095",
    journal = "JHEP",
    volume = "12",
    pages = "095",
    year = "2019"
}

@article{Hayashi:2021kro,
    author = "Hayashi, Takumi and Kamada, Kohei and Oshita, Naritaka and Yokoyama, Jun'ichi",
    title = "{Vacuum decay in the Lorentzian path integral}",
    eprint = "2112.09284",
    archivePrefix = "arXiv",
    primaryClass = "hep-th",
    reportNumber = "RESCEU-24/21, RIKEN-iTHEMS-Report-21",
    doi = "10.1088/1475-7516/2022/05/041",
    journal = "JCAP",
    volume = "05",
    number = "05",
    pages = "041",
    year = "2022"
}

@article{Nishimura:2023dky,
    author = "Nishimura, Jun and Sakai, Katsuta and Yosprakob, Atis",
    title = "{A new picture of quantum tunneling in the real-time path integral from Lefschetz thimble calculations}",
    eprint = "2307.11199",
    archivePrefix = "arXiv",
    primaryClass = "hep-th",
    reportNumber = "KEK-TH-2538",
    doi = "10.1007/JHEP09(2023)110",
    journal = "JHEP",
    volume = "09",
    pages = "110",
    year = "2023"
}

@article{Kaya:2018jdo,
    author = "Kaya, Ali",
    title = "{On $i\epsilon$ Prescription in Cosmology}",
    eprint = "1810.12324",
    archivePrefix = "arXiv",
    primaryClass = "gr-qc",
    doi = "10.1088/1475-7516/2019/04/002",
    journal = "JCAP",
    volume = "04",
    pages = "002",
    year = "2019"
}

@article{Brezin:1976vw,
    author = "Brezin, E. and Le Guillou, J. C. and Zinn-Justin, Jean",
    title = "{Perturbation Theory at Large Order. 1. The phi**2N Interaction}",
    reportNumber = "SACLAY-DPh-T/76-102",
    doi = "10.1103/PhysRevD.15.1544",
    journal = "Phys. Rev. D",
    volume = "15",
    pages = "1544--1557",
    year = "1977"
}

@article{Gamow:1928zz,
    author = "Gamow, G.",
    title = "{Zur Quantentheorie des Atomkernes}",
    doi = "10.1007/BF01343196",
    journal = "Z. Phys.",
    volume = "51",
    pages = "204--212",
    year = "1928"
}

@article{Siegert:1939zz,
    author = "Siegert, A. J. F.",
    title = "{On the Derivation of the Dispersion Formula for Nuclear Reactions}",
    doi = "10.1103/PhysRev.56.750",
    journal = "Phys. Rev.",
    volume = "56",
    pages = "750--752",
    year = "1939"
}

\end{document}